%% file: AuMicHarps.tex
\documentclass[fleqn,usenatbib]{mnras}

\usepackage{newtxtext,newtxmath}

\usepackage[T1]{fontenc}

\DeclareRobustCommand{\VAN}[3]{#2}
\let\VANthebibliography\thebibliography
\def\thebibliography{\DeclareRobustCommand{\VAN}[3]{##3}\VANthebibliography}

\usepackage{enumitem}
\usepackage{graphicx}	
\usepackage{amsmath}	
\usepackage{xcolor}
\usepackage{xspace}

\usepackage{bm}		      
\usepackage{pdflscape}	  
\usepackage{threeparttable}
\usepackage{fontawesome}  

\newcommand{\pyaneti}{\href{https://github.com/oscaribv/pyaneti}{\texttt{pyaneti}\,\faGithub}}
\newcommand{\mesaplanet}{\href{https://github.com/jo276/MESAplanet}{\texttt{MESAplanet}\,\faGithub}}

\newcommand{\citlalatonac}{\texttt{citlalatonac}}

\newcommand{\sshk}{$S_{\rm HK}$}

\newcommand{\halpha}{$\rm H_{\alpha}$}

\newcommand{\lbe}{$\lambda_{\rm e}$}
\newcommand{\lbp}{$\lambda_{\rm p}$}
\newcommand{\pgp}{$P_{\rm GP}$}

\newcommand{\ms}{${\rm m\,s^{-1}}$}
\newcommand{\mmss}{${\rm m^2\,s^{-2}}$}
\newcommand{\kms}{${\rm km\,s^{-1}}$}

\newcommand{\vsini}{$v \sin i$}
\newcommand{\logg}{$\log g$}

\newcommand{\ktwo}{\emph{K2}}
\newcommand{\tess}{\emph{TESS}}

\newcommand{\serval}{\texttt{SERVAL}}
\newcommand{\terra}{\texttt{HARPS-TERRA}}
\newcommand{\astropy}{\texttt{astropy}}
\newcommand{\cheops}{\emph{CHEOPS}}

\newcommand{\mstar}{$M_{\rm \star}$\xspace}
\newcommand{\rstar}{$R_{\rm \star}$\xspace}

\newcommand{\teff}{$T_{\mathrm{eff}}$\xspace}

\newcommand{\Prot}{$P_{\mathrm{rot}}$\xspace}
\newcommand{\iorb}{$i_{\mathrm{orb}}$\xspace}

\newcommand{\target}{AU\,Mic}
\newcommand{\targetb}{AU\,Mic\,b}
\newcommand{\targetc}{AU\,Mic\,c}

\input{aumic_params}

\input{model1_drs}             
\input{model2_bis}             
\input{aumic_pyaneti_params}   

\defcitealias{Plavchan2020}{P20}
\defcitealias{Klein2021}{K21}
\defcitealias{Martioli2021}{M21}
\defcitealias{Cale2021}{C21}
\defcitealias{pyaneti2}{B22}

\title[HARPS AU Mic planet masses]{One year of AU Mic with HARPS: I - measuring the masses of the two transiting planets}

\author[Zicher et al.]{
Norbert Zicher$^{1}$\thanks{E-mail: norbert.zicher@physics.ox.ac.uk},
Oscar Barrag\'an$^{1}$,
Baptiste Klein$^{1}$,
Suzanne Aigrain$^{1}$,
James E. Owen$^{2}$,              
Davide Gandolfi$^{3}$,         
\newauthor
Anne-Marie Lagrange$^{4}$,     
Luisa Maria Serrano$^{3}$,     
Laurel Kaye$^{1}$,
Louise Dyregaard Nielsen$^{1,5}$,
\newauthor
Vinesh Maguire Rajpaul$^{6}$,
Antoine Grandjean$^{4}$,       
Elisa Goffo$^{3,7}$,           
Belinda Nicholson$^{1,8}$
\\
$^{1}$ Sub-department of Astrophysics, Department of Physics, University of Oxford, Oxford, OX1 3RH, UK \\
$^{2}$ Astrophysics Group, Blackett Laboratory, Imperial College London, London, SW7 2AZ, U.K \\
$^{3}$ Dipartimento di Fisica, Universit\`a degli Studi di Torino, Via Pietro Giuria 1, 10125 Torino, Italy  \\
$^{4}$ Univ. Grenoble Alpes, CNRS, IPAG, 38000 Grenoble, France \\
$^{5}$ Geneva Observatory, University of Geneva, Chemin des Mailettes 51, 1290 Versoix, Switzerland \\
$^{6}$ Astrophysics Group, Cavendish Laboratory, University of Cambridge, J. J. Thomson Avenue, Cambridge, CB3 0HE, UK \\
$^{7}$ Thüringer Landessternwarte Tautenburg, Sternwarte 5, D-07778 Tautenburg, Germany \\
$^{8}$ University of Southern Queensland, Centre for Astrophysics, Toowoomba, Australia
}

\date{Accepted XXX. Received YYY; in original form ZZZ}

\pubyear{2021}

\begin{document}
\label{firstpage}
\pagerange{\pageref{firstpage}--\pageref{lastpage}}
\maketitle

\begin{abstract}
The system of two transiting Neptune-sized planets around the bright, young M-dwarf \target\ provides a unique opportunity to test models of planet formation, early evolution, and star-planet interaction. However, the intense magnetic activity of the host star makes measuring the masses of the planets via the radial velocity (RV) method very challenging. We report on a 1-year, intensive monitoring campaign of the system using 91 observations with the HARPS spectrograph, allowing for detailed modelling of the $\sim 600$\,\ms\ peak-to-peak activity-induced RV variations.
We used a multidimensional Gaussian Process framework to model these and the planetary signals simultaneously. We detect the latter with semi-amplitudes of $K_{\rm b}=$ \kb\ and $K_{\rm c}=$ \kc, respectively. The resulting mass estimates, $M_{\rm b}=$ \mpb\ and $M_{\rm c}=$ \mpc, suggest that planet b might be less dense, and planet c considerably denser, than previously thought. These results are in tension with the current standard models of core-accretion. They suggest that both planets accreted a H/He envelope that is smaller than expected, and the trend between the two planets' envelope fractions is the opposite of what is predicted by theory. 

\end{abstract}



\section{Introduction}

Planets orbiting young stars offer a unique window into the formation and evolution of planets and planetary systems. The first few hundreds of Myr, when the planets evolve most rapidly, and their observable parameters are still affected by initial conditions, offer the most sensitive tests of theoretical models. Planets whose masses and radii can be measured directly are particularly valuable in that respect, but only a few are known to date. This is in large part due to the rapid rotation and enhanced magnetic activity of the host stars, which hinders both transit detection and radial velocity (RV) follow-up. 

In recent years, the  \ktwo\ \citep[][]{Howell2014} and \tess\ \citep[][]{Ricker2015} space missions have enabled the detection of a number of planets transiting young stars in nearby open clusters and associations, including \citep{David2016a,David2016b,David2019,Mann2016a,Mann2016b,Mann2017,Mann2018,Mann2020,Gaidos2017,Pepper2017,Vanderburg2018,Rizzuto2018,Rizzuto2020,Newton2019,Newton2021, Kossakowski2021}, but most of them currently lack mass determinations. Dedicated spectroscopic surveys have also uncovered a number of non-transiting planets around very young stars (including \citep{Quinn2012,Donati2016,yu2017}), but these lack radius estimates. The combination of measured radius and mass gives access to the planet's mean density, which in turn contains clues regarding its bulk and atmospheric composition. Specifically, young planets ($\lesssim 100$~Myr) with orbital periods $\lesssim 30~$days and known mean densities are particularly valuable, as this is the period range where planets are most strongly effected by atmospheric mass-loss over their lifetimes \citep[e.g.][]{Lopez2013,Owen2019}. Therefore, young planets provide the opportunity to probe the amount of primordial gas these planets accreted from their parent protoplanetary discs before the majority of it is lost. In addition, for those planets that are either close enough to their star, or have low enough mass that mass-loss is important on timescales comparable to their age then constraints can be placed on their initial entropy \citep{Owen2020}. Knowledge of the amount of primordial gas these planets accreted, and in what thermodynamic state, is extremely valuable as the origin of close-in planets remains uncertain \citep[e.g.][]{Bean2021}.  To date, there are only a few planets with well-determined masses \emph{and} radii orbiting stars with well-determined ages below 1\,Gyr \citep{Mann2017,David2019b,Barragan2019,Barragan2021b,Plavchan2020,Klein2021,Kossakowski2021,suarez2021}.

Two of them, K2-100b \citep{Mann2017,Barragan2019} and TOI-1201b \citep{Kossakowski2021}, orbit stars with ages in the range 600--800\,Myr. K2-100b is 
a $3.88\pm 0.16\,R_\oplus$, $21.0\pm 6.2\,M_\oplus$ planet on a $1.67$\,d orbit around an early G-dwarf in the $\sim 750$\,Myr-old Praesepe open cluster, whose transits were discovered in data from the \ktwo\ space mission \citep{Howell2014}, and whose mass was measured using HARPS-North \citep{Pepe2010}. Photo-evaporation models suggest that K2-100b is likely still losing its atmosphere \citep[see][and references therein]{Barragan2019}, but any estimate of both its original and its final composition depends critically on assumptions regarding the UV flux of the host star and how it has evolved throughout its lifetime. TOI-1201b is 
a $2.42\pm 0.09\,R_\oplus$, $6.3\pm 0.9\,M_\oplus$ planet on a $2.49$\,d orbit around an early M-dwarf whose gyrochronological age lies in the range 600--800\,Myr and which is a likely kinematic member of the Hyades super-cluster. Its transits were discovered in data from the \tess\ space mission \citep{Ricker2015}, and its mass was measured using CARMENES \citep{Quirrenbach2014}. Both host stars are relatively faint in the optical, so additional follow-up to refine the planets' masses or detect their atmosphere is best pursued with infrared instruments.

By contrast, another of these planets orbits \target, the brightest M-dwarf in the Southern sky (V = 8.73 \citealt{Torres2006}). 
\citet[][hereafter \citetalias{Plavchan2020}]{Plavchan2020} discovered the transits of \targetb\ in the first month of data from the Transiting Exoplanet Survey Satellite (TESS, \citealt{Ricker2015}), reporting an orbital period of $8.46$\,d and a planetary radius of $4.2\pm 0.2\,R_\oplus$. Due to its young age \citep[$22\pm3$\,Myr;][]{Mamajek_Bell2014}, the \target\ system provides a much more direct test of planet formation and evolution models, but its intense magnetic activity also exacerbates the challenges of detecting both the transits and the RV signal of any planetary companions. In addition, \target\ hosts an edge-on debris disc \citep[e.g.,][]{kalas2004,Boccaletti2015}, which shows peculiar fast moving features, still unexplained \citep{Boccaletti2018}. The \tess\ light curve exhibits significant variability ($\sim$0.1\,mag) due to starspots, from which \citetalias{Plavchan2020} determined the star's rotation period, $P_{\rm rot}=4.863$\,d, as well as frequent flares, which hinder the detection and modelling of the transits. \citetalias{Plavchan2020} combined data from several optical and near-infrared spectrographs, obtained over several years (starting long before the discovery of the transits) in an effort to constrain the planet's mass, but obtained only an upper limit. A more intensive observing campaign using the SPIRou near-infrared spectropolarimeter \citep{Donati2020}, focusing on several consecutive rotational cycles, enabled \citet[][hereafter \citetalias{Klein2021}]{Klein2021} to measure a mass of $17.1^{+4.7}_{-4.5}\,M_\oplus$ for \targetb, recently refined at $20.1^{+1.7}_{-1.6}\,M_\oplus$ in \citet{Cale2021}. 

\citetalias{Plavchan2020} also reported an additional transit-like event from a possible second transiting planet. \tess\ re-observed the system two years later, enabling \citet[][hereafter \citetalias{Martioli2021}]{Martioli2021} to confirm the detection of \targetc, with a radius of $3.24\pm 0.16\,R_\oplus$ planet and a period of $18.859$\,d, though no more than a 5$\sigma$ upper limit of $20.13\,M_\oplus$ has been reported for its mass \citep{Cale2021}. Note that the existence of more than one transiting planet around \target\ is not altogether surprising, given that \target\ hosts an edge-on debris disc, and that observations of the transits of \targetb\ have shown its orbit to be aligned \citep[including][]{Addison2021, Martioli2020, Szabo2021, Hirano2020}. The parameters of \target\ and its two transiting planets that are relevant in the present work, as reported in the literature at the time of writing, are listed in Table~\ref{tab:literature}. 

\begin{table*}
\begin{center}
  \caption{Parameters of the \target host star and transiting planets gathered from the literature. \label{tab:literature}}  
  \begin{tabular}{l c c}
    \hline
    \hline
    \noalign{\smallskip}
    Parameter & Value & Reference/Notes \\
    \noalign{\smallskip}
    \hline
    \noalign{\smallskip}
    \multicolumn{3}{l}{\emph{\bf Au\,Mic\ stellar parameters }} \\
    \noalign{\smallskip}
        Distance & \sdistance & \citet{GaiaEDR3} \\
        \teff & \stemp & \citet{Afram_Berdyugina2019} \\
         Radius (\rstar) & \sradius & \citet{White2019} \\
         Mass (\mstar) & \smass & \citetalias{Plavchan2020} \\
        \logg & \slogg & computed from \rstar \& \mstar \\
        Luminosity & \slum & \citet{Plavchan2009} \\
        Age & \sage & \citet{Mamajek_Bell2014} \\
        \Prot & \srot & \citetalias{Plavchan2020} \\
        \iorb & \siorb & \citetalias{Plavchan2020} \\
        \vsini & \svsini & \citetalias{Klein2021} \\
        Linear limb dark. coef & \sllimb & \citet{Claret2018} \\
        Quadratic limb dark. coef & \sqlimb & \citet{Claret2018} \\
    \noalign{\smallskip}
    \hline
    \multicolumn{3}{l}{\emph{\bf Au\,Mic\,b}} \\
    \noalign{\smallskip}
        Transit epoch $T_0$ (BJD$_\mathrm{TDB}-$2\,450\,000)  & \Tzeroblit & \citetalias{Martioli2021} \\ 
        Period & \pblit & \citetalias{Martioli2021} \\ 
        Semi-major axis & \semib & Kepler's law \\
        Impact parameter & \impactb & \citetalias{Martioli2021} \\
        Planet radius & \radiusblit & \citetalias{Martioli2021}\\
        Velocity semi-amplitude & \kblit & \citetalias{Klein2021} \\
        Planet Mass & \massblit & \citetalias{Klein2021} \\
        Equilibrium temperature & \eqtempb & \citetalias{Martioli2021} \\
    
    \noalign{\smallskip}
    \hline
    \multicolumn{3}{l}{\emph{ \bf Au\,Mic\,c}} \\
    \noalign{\smallskip}
        Transit epoch $T_0$ (BJD$_\mathrm{TDB}-$2\,450\,000)  & \Tzeroclit & \citetalias{Martioli2021} \\
        Period & \pclit & \citetalias{Martioli2021} \\
        Semi-major axis & \semic & Kepler's law \\
        Impact parameter & \impactc & \citetalias{Martioli2021} \\
        Planet radius & \radiusclit & \citetalias{Martioli2021} \\
        Velocity semi-amplitude & \kclit & \citetalias{Klein2021} \\
        Planet Mass & \massclit & \citetalias{Martioli2021} \\
        Equilibrium temperature & \eqtempc & \citetalias{Martioli2021} \\
    
    \hline
  \end{tabular}
\end{center}
\end{table*}

Together with \target's youth and proximity, the fact that it hosts not one but two transiting planets opens up the possibility of comparative exoplanetology. It is thus the target of a number of ongoing ground- and space-based observing campaigns aiming to measure the planets' masses more precisely, as well as to detect the planet's atmosphere and exosphere in transmission\footnote{See e.g.\ HST Programme 15836, PI E.\ Newton.}. Knowing the planet's mass is important not only to understand its bulk composition, but also to interpret any transmission spectrum observations, as it helps break degeneracies in the atmospheric retrieval process \citep{Batalha2019}. 

In this work, we report on the results of a 10-month intensive monitoring campaign on \target\ with HARPS, which aims to characterise the activity-induced RV signal of the host star sufficiently to detect the two transiting planets. We use HARPS because activity signals are well characterised in the optical, where they are significantly strong signals.
As our approach relies on modelling the effects of stellar activity alongside the planetary signal, rather than filtering one to reveal the other, working in a regime where activity signals are strong can be an advantage rather than an impediment.

This manuscript is part of a series of papers under the project \emph{GPRV: Overcoming stellar activity in radial velocity planet searches} funded by the European Research Council (ERC, P.I.~S.~Aigrain), and will be followed by two companion papers (namely Klein et al., submitted, and Zicher et al., in prep.). This paper is structured as follows. Section~\ref{sec:obs} gives details of the observations and data reduction, as well as the identification of epochs affected by stellar flares, which were discarded from the rest of the analysis. We describe the framework we use to model the RVs and activity indicators in Section~\ref{sec:multigp}, where we detail the simulations done, using the same framework, to optimize the observing strategy ahead of time. We discuss our results and confront them with theoretical models of early planet evolution in Section~\ref{sec:discuss}. Finally, we conclude and discuss future prospects for the characterisation of planetary signals orders of magnitude below the stellar noise in Section~\ref{sec:concl}. 

\section{RV observations}
\label{sec:obs}

AU Mic has been monitored by a number of optical and near-IR RV spectrographs over the past decade, both before and after the discovery of the transiting planets, including: HARPS, iShell, HIRES, CARMENES, SPIRou, CHIRON, IRD, TRES, Minerva Australis \citep[\citetalias{Plavchan2020,Klein2021};][hereafter \citetalias{Cale2021}]{Cale2021}. However, none of the individual datasets so far have sufficient time-sampling to constrain both the activity signals and the orbits of the transiting planets on their own. Combining data from multiple instruments is possible, and can in principle be advantageous, if the chromaticity of the activity signal is explicitly accounted for when modelling the data \citepalias{Cale2021}. However, combining data from multiple instruments with different wavelength ranges is also challenging because each instrument has different zero-point and noise characteristics, as well as qualitatively distinct activity indicators. Furthermore, the sparse time-sampling of many of these datasets compared to the rotation period of AU Mic, as well as the orbital period of both planets, makes them of limited use to constrain the planetary masses and orbits. The datasets with the best time-sampling to date were those presented in C21, but even in that study, the largest number of nights on which \target\ was observed by a given instrument in a given season was about 40. We therefore opted to carry out a dedicated, intensive survey using a single instrument, namely HARPS, and focused exclusively on this new dataset in the present work. We defer any attempt to combine the new data with archival datasets to future work. 

\subsection{Previous HARPS observations}

The system has been observed since late $2013$ with HARPS (under programmes 192.C-0224, 098.C-0739, 099.C-0205, 0104.C-0418, PI Lagrange), as part of a wider, long-term monitoring survey aiming to detect Jovian planets at intermediate separations \citep{Grandjean2020} around targets of the SPHERE direct imaging survey \citep{Beuzit2019}. In total there are 54 archival spectra taken on 29 individual nights between 2 October 2013 and 2 November 2019. As these observations were primarily intended to search for giant planets beyond the snow line, their time sampling is too sparse to constrain the orbits of the transiting planets effectively, but we extracted the RVs and activity indicators from the ESO archive when planning our own programme, in order to evaluate the magnitude of the activity-induced RV variations and to estimate the number of new observations required. We note that neither this HARPS programme, nor the SPHERE survey have reported any planet detections at wide separations around \target\ so far \citep{Lannier2017}. In future work, we plan to combine both new and archival HARPS datasets to place limits on the presence of additional planets in the system (Zicher et al. in prep.).

\input{RV_table}

\subsection{New HARPS observations}

The new HARPS observations were collected during ESO Periods 106 \& 107 under programmes 0105.C-0288\footnote{The P105 observations were delayed because of the COVID-19 pandemic and taken in P107.} \& 0106.C-0852 (PIs Aigrain and Zicher). In total, we obtained 91 individual spectra: 22 observations on 20 individual nights in P106 (between 15 November and 9 December 2020), and 69 observations on 49 individual nights in P107 (between 24 May and 22 September 2021). All observations were taken in High-Accuracy Mode (HAM), with resolution $R=115,000$ \citep{Mayor2003}. The exposure time was initially set to $900$s, but increased to $1200$s as the seeing became more variable from June 2021 (onset of Chilean winter), resulting in a typical signal to noise ratio (SNR) at $550\ \rm nm$ of $80$ to $100$. Both programmes were carried out under the auspices of the HARPS time-share programme organised by F.\ Bouchy and X.\ Dumusque.

\subsection{Data reduction and time-series extraction}

\begin{figure*}
    \centering
    \includegraphics[width=\linewidth]{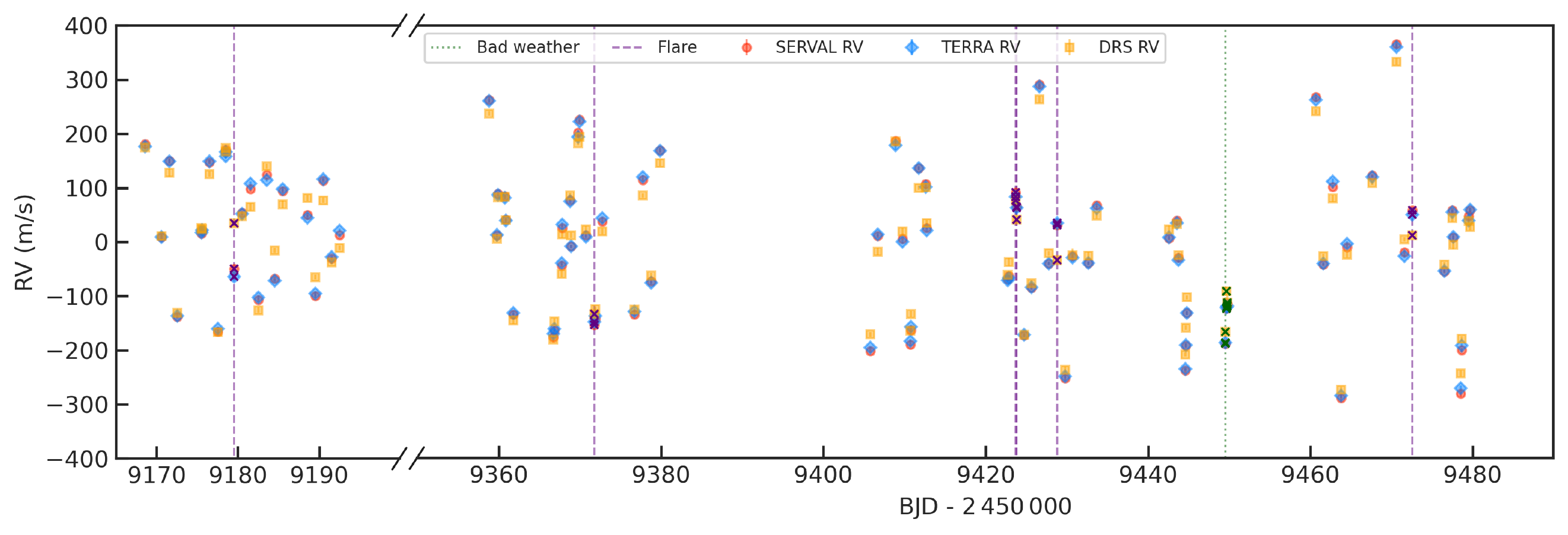}
    \caption{Mean-subtracted RV time-series as extracted with SERVAL (red circles), TERRA (blue diamonds), and the DRS (yellow squares).
    Note the gap in the $x$-axis corresponding to a $\sim 5$-month break in the observations. The vertical lines show the observations affected by flares (purple dashed) and clouds (green dotted), which are also marked by `x' symbols, and were discarded from the rest of the analysis (see Section~\ref{sec:outliers} for details).}
    \label{fig:data-timeseries}
\end{figure*}

The data were reduced using version $3.8$ of HARPS Data Reduction Software (DRS). DRS RVs were extracted from the Cross-Correlation Function (CCF) obtained using an M4 mask. We also extracted RVs using two alternative pipelines which have been shown to outperform the DRS for M-dwarfs: \serval\ \citep{SERVAL} and \terra\ \citep{TERRA}. \serval\ and \terra\ both compute a template spectrum from the observations themselves, rather than using a pre-existing digitized mask to compute the CCF as in the DRS. In the case of \serval, the template is constructed by co-adding the observed spectra in the barycentric rest-frame. In the case of \terra, the template is initially computed by co-adding high SNR observations, allowing preliminary estimates of the RVs to be made, then a new, higher SNR template is produced by co-adding all the spectra in the stellar rest frame, and the RVs are evaluated again. The median RV uncertainties for the DRS, \serval\ and \terra\ were 5.2, 3.6 and 3.4\,\ms, respectively, while the Root Mean Square (RMS) RV variations are 121, 130 and 132\,\ms, respectively. Even though all three time-series appear qualitatively similar (see Figure~\ref{fig:data-timeseries}), we observe a much better agreement between \serval\ and \terra\ RVs (median difference of $\sim$5\,\ms), than between either and the DRS (median difference between \terra\ and DRS RVs $\sim$20\,\ms). As outlined above, this is not surprising, as \terra\ and \serval\ both construct a template from the observed spectra, whereas the DRS M4 mask is optimised for older, less active and somewhat cooler stars than \target.

\begin{figure*}
    \centering
    \includegraphics[width=0.99\textwidth]{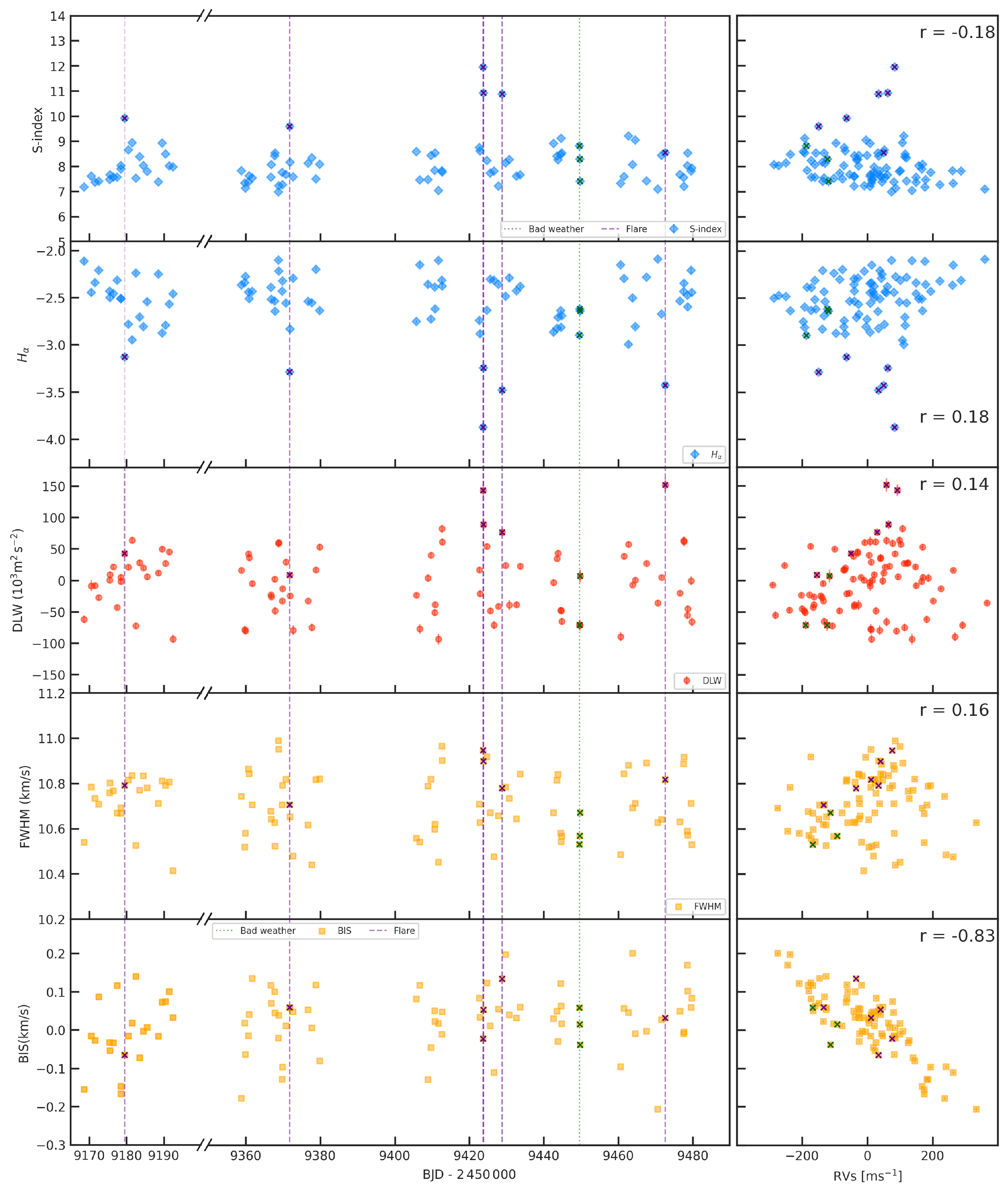}
    \caption{Activity indicators versus time (left) and versus the RVs produced by the corresponding pipeline (right). From top to bottom: DRS FWHM and BIS, \serval\ DLW, \terra\ \sshk\ and H$\alpha$ index. The colour indicates the pipeline used to extract each indicator: blue for \terra\, red for \serval\ and yellow for the DRS. Significant outliers in either \sshk\ or \halpha, which indicate that the corresponding observations were affected by flaring, are marked by purple dashed vertical lines in the left column. Along with one observation affected by clouds (green dotted vertical line), these observations (marked as `x' symbols in all panels) were discarded from the rest of the analysis (see Section~\ref{sec:outliers} for details). }
    \label{fig:data-actind}
\end{figure*}

We also extracted a number of ancillary time-series which might serve as activity indicators. These fall into two categories: those that quantify changes in the mean line profile, and those that measure chromospheric emission in the cores of certain spectral lines. In the former category, we include the Full Width at Half Maximum (FWHM) and Bisector Inverse Slope (BIS) of the CCF provided by the DRS, as well as the Differential Line Width (DLW) computed by the \serval\ pipeline. The DLW is obtained by correlating the fit residuals with the second derivative of the template \citep{SERVAL}. The chromospheric indicators we consider include the  Mt Wilson \sshk\ and \halpha\ indices computed by the \terra\ pipeline\footnote{Note that the chromospheric indicators provided by the \serval\ pipeline and the DRS are entirely consistent with those provided by \terra, with Pearson correlation coefficients systematically larger than 0.99.}, which measure emission in the cores of the CA {\sc ii} H \& K and the \halpha\ lines, respectively. As presented in Sections~\ref{sec:outliers} and \ref{sec:choice_act}, the chromospheric activity indicators were used to identify observations affected by flares, while the line-shape indicators were most useful to model the activity-induced signals in the RVs. 

All the HARPS RVs and activity indicators after June 2015 (after the HARPS fibre upgrade \citep {LoCurto2015} are given in Table~\ref{tab:rvs} - the full version of which is available in machine-readable format in the supplementary online material. The activity indicator time-series are shown in Figure~\ref{fig:data-actind}. 

\subsection{Removing flares and clouds}
\label{sec:outliers}

As well as displaying significant rotational modulation of starpots, the ground-based and \tess\ light curves of \target\ show frequent white-light flares (e.g., \citealt{Hebb2007}; \citetalias{Plavchan2020,Martioli2021}). These flares can significantly affect the measured radial velocities as they distort the line profiles. As our current models are not equipped to account for the RV effects of flares, we flagged the observations affected by flares and discarded them from the analysis. The procedure used to identify observations strongly affected by flares is described in detail in the companion paper Klein et al. (submitted). In short, we computed a chromospheric emission metric for 7 different chromospheric lines, namely Ca\,II H \& K (resp. 3968.47 \& 3933.66 {\AA}), H$\alpha$ (6562.808 {\AA}), H$\beta$ (4861.363 {\AA}), Na\,I D1 \& D2 (resp. 5895.92 \& 5889.95 {\AA}) and He\,I D3 (5875.62\,{\AA}) using the method of \citet{SERVAL} and the integrations windows of \citet{gomes2011}. As the effects of stellar flares might change from one chromospheric line to the other, we defined a master index by taking the median-normalized average of all chromospheric emission indices. We finally applied a a 3$\sigma$-clipping process to the resulting time-series,
flagging a total of 6 observations. These were deemed likely to be affected by a stellar flare, and were removed from the subsequent analysis (see Figure~\ref{fig:data-actind}). Note that the RVs affected by flares do not stand out as outliers in the raw RVs, but would do so after subtracting the activity model described in Section~\ref{sec:multigp}.


After removing the observations affected by flares, one substantial outlier ($4-\sigma$) remained when looking at the residual time-series (after subtraction of the best-fit stellar activity plus planetary signals model, as described in Section~\ref{sec:modellingrvs}). We inspected images from the Danish all-sky camera at La Silla, which showed that variable, high cirrus were present that entire night. We then inspected all three observations of AU Mic taken that night, and found the spectra to be strongly contaminated by the Moon. We thus removed them from further analysis. Although only one of the three observations taken that night was a significant RV outlier, the spectra were all affected by the Moon so we deemed it safer to remove them all. We checked that no other spectra taken on other nights were affected in the same way by dividing each spectrum by the median spectrum across all observations. The three observations from that night were the only ones which showed strong evidence of Moon contamination.

All the observations affected by flares or clouds are indicated in Table~\ref{tab:rvs}. The total number of epochs remaining and included in our final analysis, shown in Figure~\ref{fig:data-timeseries}, was therefore 82. 

\subsection{Correlations and periodograms}

\begin{figure*}
    \centering
    \includegraphics[width=\linewidth]{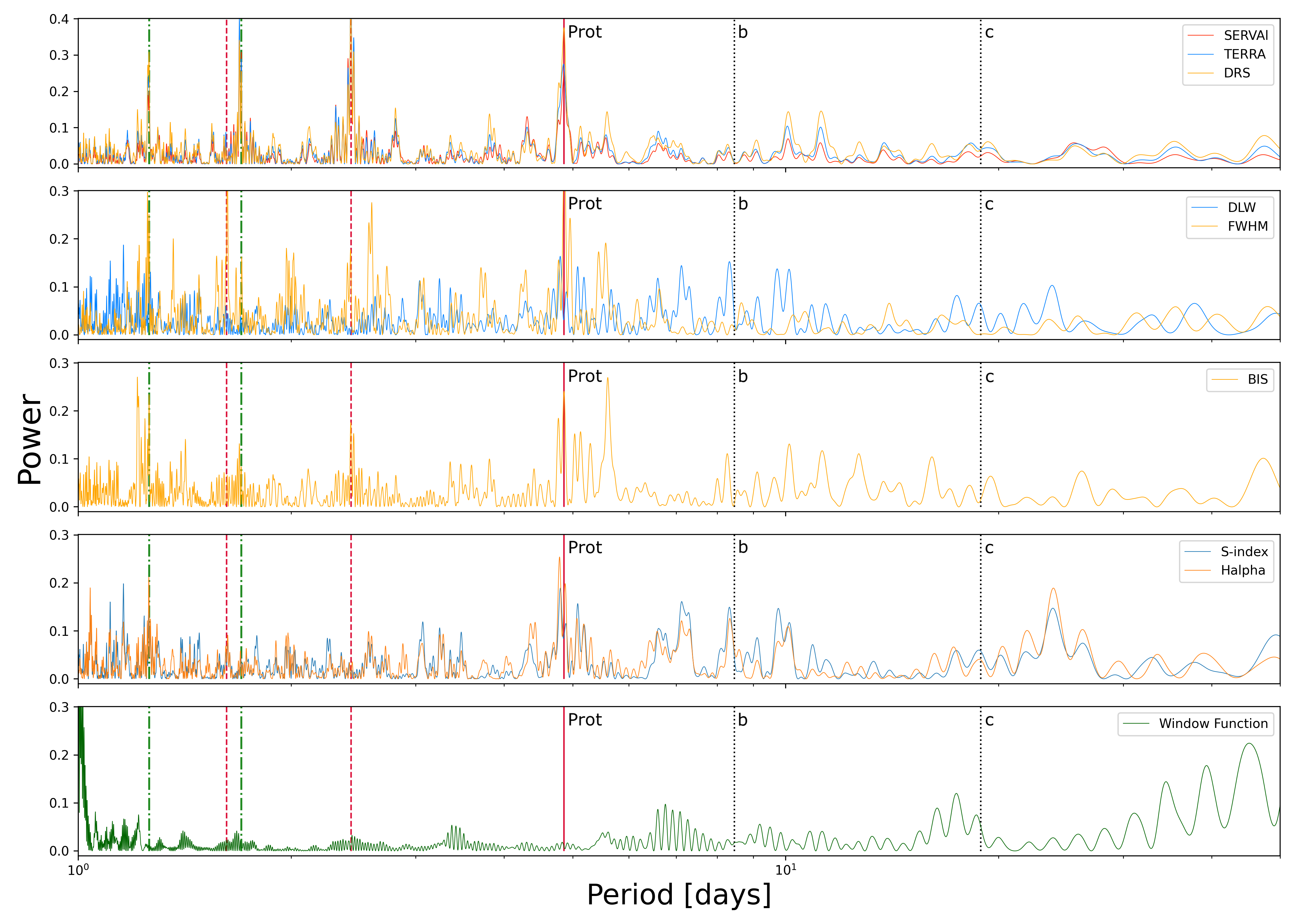}
    \caption{Lomb-Scargle periodograms of the 3 RVs time-series (top), DLW \& FWHM ($2^{\rm nd}$ panel), BIS ($3^{\rm rd}$ panel), S-index \& \halpha ($4^{\rm th}$ panel) and the window function of the observations (bottom). Vertical crimson lines denote the stellar rotation period and its first two harmonics, vertical green lines denote 1-d aliases of the stellar rotation period, while vertical black dotted lines denote the orbital periods of \targetb\ and c.}
    \label{fig:lsper}
\end{figure*}

In the right-hand column of Figure~\ref{fig:data-actind}, we show the correlation between each activity indicator and the RVs extracted using the corresponding pipeline. Each plot is annotated with the corresponding Pearson correlation coefficient. Although all the time-series are dominated by activity signals, this does not result in a clear correlation between the activity indicators and the RVs, with the exception of the BIS, which is anti-correlated with the RVs. This is fully expected, as the other activity indicators depend only on the projected area of the active regions, their contrasts, and (for FWHM and DLW) on the absolute value of the surface radial velocity at the location of the active regions relative to the stellar rest frame, but the RVs and BIS also depend on the sign of this quantity.

Lomb-Scargle periodograms of the RVs and activity indicators time-series are shown in Figure~\ref{fig:lsper}, along with that of the observation window function. Both RV and DLW time-series show strong peaks at the rotation period and its first harmonic. There is also a peak in both periodograms near, but not exactly at, the second harmonic ($P_{\rm rot}/3$). Peak splitting around the rotation period and its harmonics can indicate differential rotation or active region evolution (see e.g., \citealt{Aigrain2012}; \citetalias{Klein2021}). The raw RVs do not display significant peaks at the orbital period of either planet. There are no prominent peaks in the window function, although there is a little excess power around, but not exactly at, the period of \targetc. These periodograms were calculated using the {\astropy} package \citep{astropy2013,astropy2018}.

\section{Modelling the RVs}
\label{sec:modellingrvs}

In this section, we describe the method used to analyse the data, which involved jointly modelling activity and planet signals in the RVs and the activity indicators simultaneously, using the multidimensional Gaussian Process (GP) model introduced by \citet{Rajpaul2015} as implemented in the \pyaneti\ package \citep[][hereafter \citetalias{pyaneti2}]{pyaneti2}. 

\subsection{Activity model}
\label{sec:multigp}

We model the activity signals in the RVs, $V(t)$ and one or two activity indicators, $\alpha(t)$ and optionally $\beta(t)$ as 
\begin{equation}
    \begin{aligned}
    V(t) & = & A_V G(t)  + B_V \dot{G}(t) + C_V \\
    \alpha(t) & = & A_{\rm \alpha} G(t) + C_{\rm \alpha}, \\
    \beta(t) & = & A_{\rm \beta} G(t) + B_{\rm \beta} \dot{G}(t) + C_{\rm \beta}, \\
    \label{eq:3mgp}
\end{aligned}
\end{equation}
\noindent where the function $G(t)$ is a latent variable, loosely representing the projected area of the visible hemisphere covered by active regions, and is modelled as a quasi-periodic GP with covariance function:
\begin{equation}
    \gamma(t_i,t_j) = \exp 
    \left[
    - \frac{\sin^2[\pi(t_i - t_j)/P_{\rm GP}]}{2 \lambda_{\rm P}^2}
    - \frac{(t_i - t_j)^2}{2\lambda_{\rm e}^2}
    \right],
    \label{eq:gamma}
\end{equation}
where \pgp\ is the characteristic period of the GP, and corresponds to the stellar rotation period, \lbp\ the inverse harmonic complexity, and \lbe\ is the evolution timescale, and is related to the lifetime of the active regions. The parameters $A_V$, $B_V$, $A_{\rm \alpha}$, $A_{\rm \beta}$ and $B_{\rm \beta}$ are free parameters, which control the relationship between the latent GP variable and the observables, while $C_V$, $C_{\rm \alpha}$ and $C_{\rm \beta}$ are constant offsets for each time-series, which are also free parameters. The interested reader is referred to \citet{Rajpaul2015} and \citetalias{pyaneti2} for a full description of the activity model.

Chromospheric activity indicators, such as \sshk\ or \halpha, depend primarily on the fraction of the visible disc that is covered in active regions, and are thus expected to be approximately proportional to $G(t)$ and were thus modelled under the form $\alpha(t)$ in Equation~(\ref{eq:3mgp}). The same is true of the FWHM and DLW, which measure the width of the spectral lines. On the other hand, the BIS, which measures the asymmetry of the spectral lines, is expected to depend on $\dot{G(t)}$, and was thus modelled like $\beta(t)$ in  Equation~(\ref{eq:3mgp}).

\subsection{Joint activity and planet model}

To model the activity signal alongside the signals of the known transiting planets, we subtract from the RVs the sum of two Keplerian signals, and compute the likelihood of the GP applied to the residuals (see Equation 7 in \citetalias{pyaneti2}). This introduces a further 6 or 10 free parameters into the model (depending on whether the orbits are assumed to be circular or not).  

The measurement uncertainties on the RVs and DLWs at each epoch were accounted for by adding a term to the diagonal of the GP covariance matrix. In addition, we add a separate `jitter' term to the diagonal of the covariance for each time-series (in effect, a constant term added in quadrature to the formal uncertainties on each observation, see Equation 13 in \citetalias{pyaneti2}). This term absorbs any imperfections in our activity plus Keplerian models and ensures that the resulting uncertainties are propagated to the final parameter estimates.

\subsection{Choice of RV version and activity indicator(s)}
\label{sec:choice_act}

We performed a number of runs using different versions of the RV time-series, and different combinations of activity indicators. We focused on the DRS FWHM \& BIS and the \serval\ DLW because their periodograms (shown in Figure~\ref{fig:lsper}) display prominent peaks at the rotation period of \target\ and its harmonics. The \sshk\ of the chromospheric activity index has a more complex periodogram, and proved to be less useful in constraining the activity signal in the RVs. For the RV time-series, we tried using both the DRS and the \serval\ versions, but we note that the \terra\ RVs would give results that are essentially identical to those obtained with the \serval\ RVs.

We tried the following models:
\begin{enumerate}[label=\arabic*), labelwidth=*]
\item $V(t)={\rm RV}({\rm DRS})$,  $\alpha(t)={\rm FWHM}$ and $\beta(t)={\rm BIS}$;
\item $V(t)={\rm RV}($\serval$)$,  $\alpha(t)={\rm DLW}$ and $\beta(t)={\rm BIS}$;
\item $V(t)={\rm RV}($\serval$)$ and  $\alpha(t)={\rm DLW}$;
\end{enumerate}
All three models give mutually consistent results, although model 1 does not yield a $3\sigma$ detection of \targetc\, whereas models 2 and 3 do. In other words, the \serval\ RVs and DLWs result in slightly better sensitivity to the planetary signals compared to the DRS RVs and FWHMs. The results of models 2 and 3 are essentially equivalent. In principle, including the BIS in the modelling should help constrain $G(t)$, but in this specific case the BIS does not seem to provide additional information that the RV and DLW time-series do not already contain. For the remainder of this paper, we therefore adopt model 3 as our fiducial model, since it is simple and has 3 fewer free parameters.
For completeness, we show the results for the three runs in Appendix~\ref{sec:threeruns}.

\subsection{Exploration of the parameter space}
\label{sec:modellingmcmc}

The full model has 20 parameters: 5 per Keplerian signal, 8 for the activity model, and 2 jitter parameters. These are listed in Table~\ref{tab:pars}, which gives the prior adopted for each parameter. The period and time of transit for each planet are tightly constrained by the \tess\ light curve and we adopted Gaussian priors based on the  ephemerides reported by \citetalias{Martioli2021}. We also adopt a uniform prior between 4.8 and 4.9 days for the GP period \pgp, based on the rotation period reported by \citetalias{Plavchan2020}. 
For the orbital eccentricities of the planets, we adopt a beta distribution prior, as advocated by \citet[][]{VanEylen2019}
for multiplanet systems. For all the other parameters, we adopted minimally informative priors.

We then explored the parameter space with a Markov chain Monte Carlo (MCMC) sampler \citep[see][for details]{pyaneti} to evaluate the joint posterior distribution over all the parameters. We iterated 250 independent Markov chains in sets of 5000 steps, checking for convergence using the auto-correlation length of the chains after each set of 5000 steps. If the chains were not converged, we iterated for a further set of 5000 steps, repeating the process until convergence was reached. We then used the last 5000 steps, thinning the chains by a factor of 10 so that samples are uncorrelated, to create the final posterior distribution (corresponding to 125\,000 independent samples for each parameter). The resulting `corner' plot, showing 1D and 2D posterior densities for all the parameters, is shown in the Appendix (Figure~\ref{fig:corner_plot}). Except for the longitude of periastron $\omega$, which is unconstrained for both planets, the marginal posterior distributions for each of the sampled parameters are uni-modal, indicating that there are no pathological degeneracies between the parameters.

\subsection{Results}
\label{sec:res}

\begin{figure*}
    \centering
    \includegraphics[width=\textwidth]{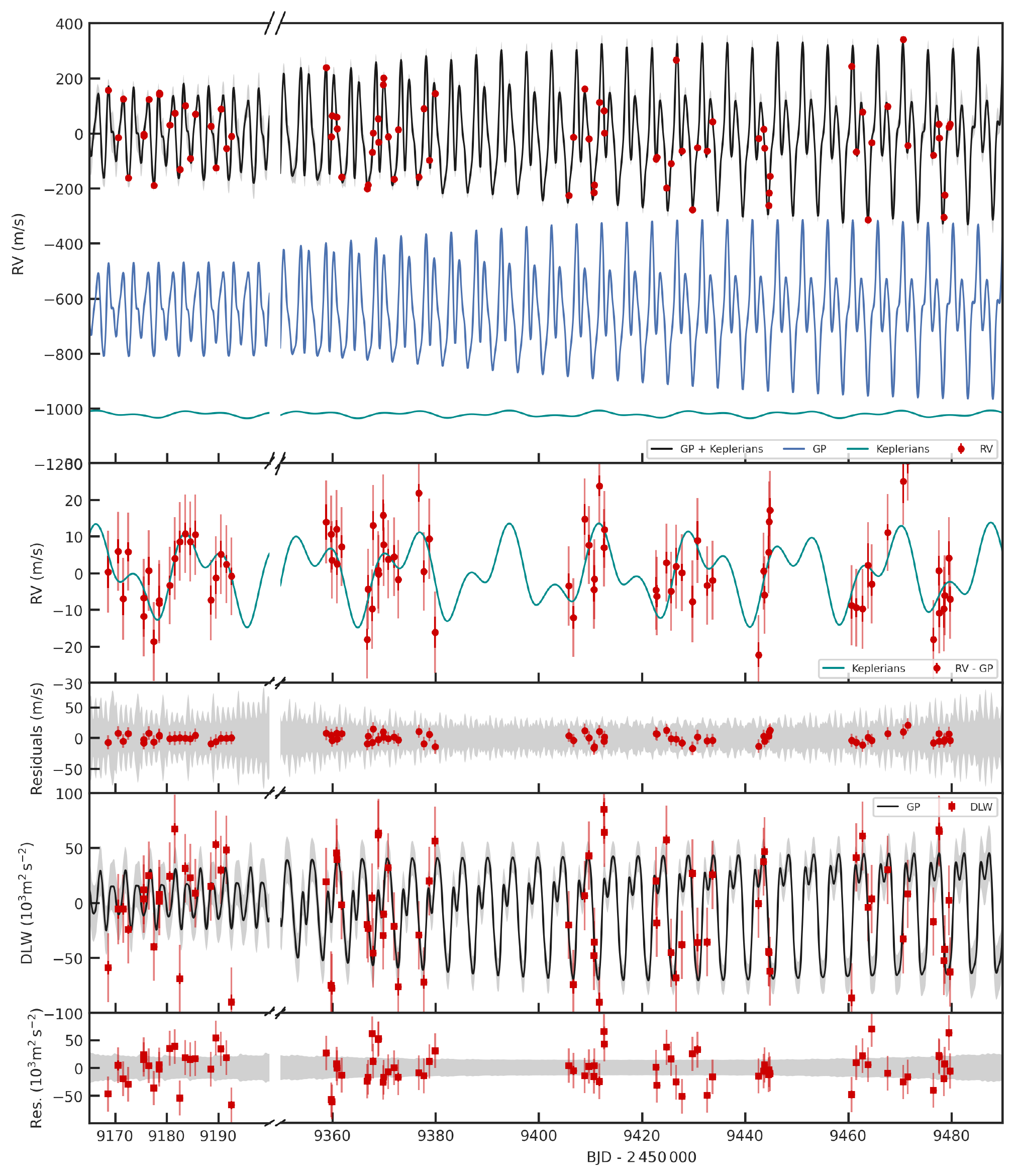}
    \caption{Radial velocity and DLW time-series (red symbols with error bars) with the best-fit model (solid lines) and $3\sigma$ confidence intervals (shaded grey areas). Note the break in the $x$-axis corresponding to a $\sim 5$-month break in the observations. The top panel shows the RV data with the full model in black. The activity and Keplerian components are also shown separately, in blue and cyan respectively, vertically offset for clarity. The second panel shows the data after subtracting the activity model, with the Keplerian component over-plotted. The final RV residuals are shown in the third panel. The fourth panel shows the DLW time-series, with the activity model in black, and the DLW residuals are shown in the fifth panel. In all the panels, the nominal error bars are in solid colour, and the error bars taking into account the jitter are semi-transparent. }
    \label{fig:timeseries}
\end{figure*}

\begin{figure*}
    \centering
    \includegraphics[width=0.49\textwidth]{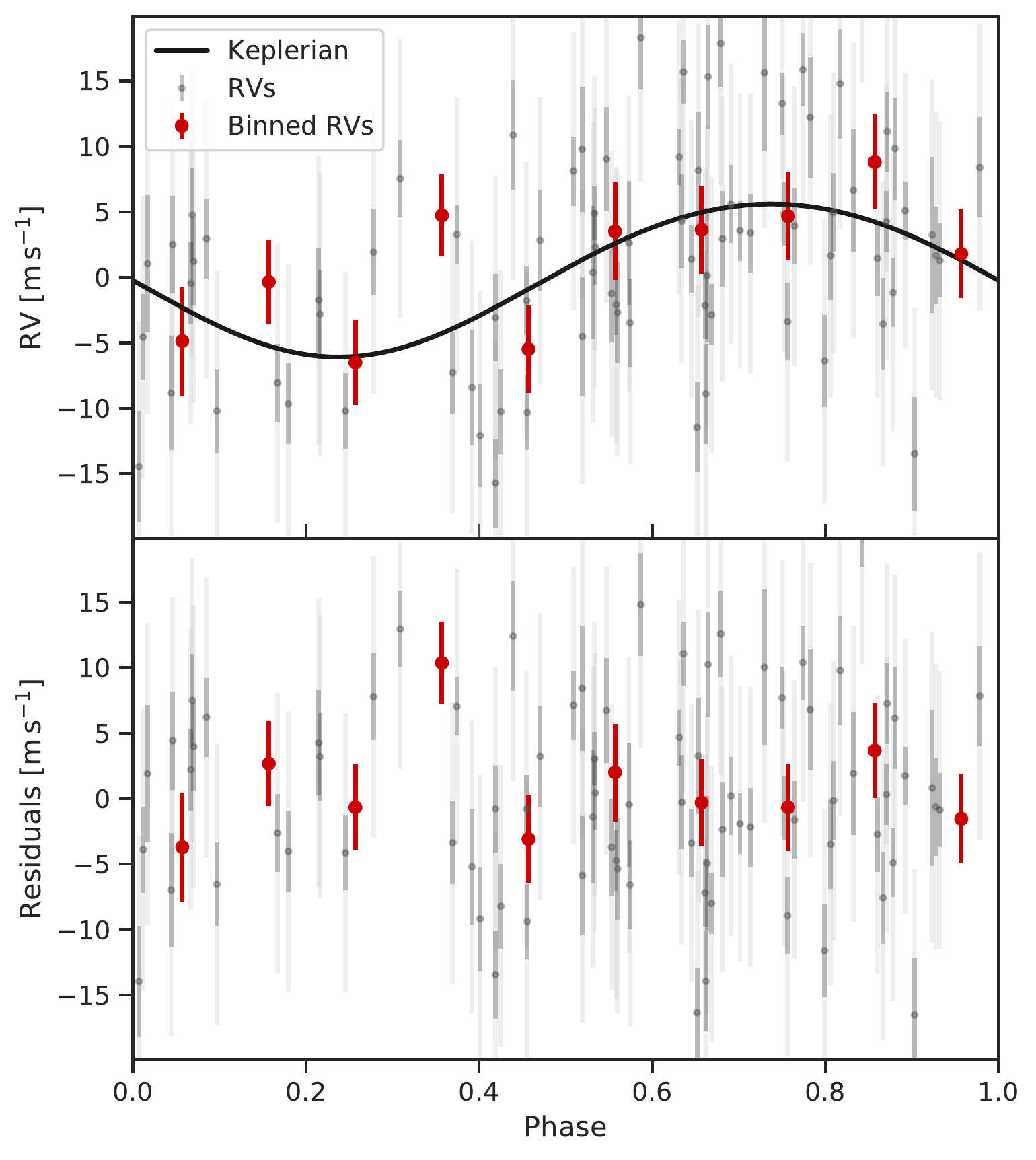}
    \includegraphics[width=0.49\textwidth]{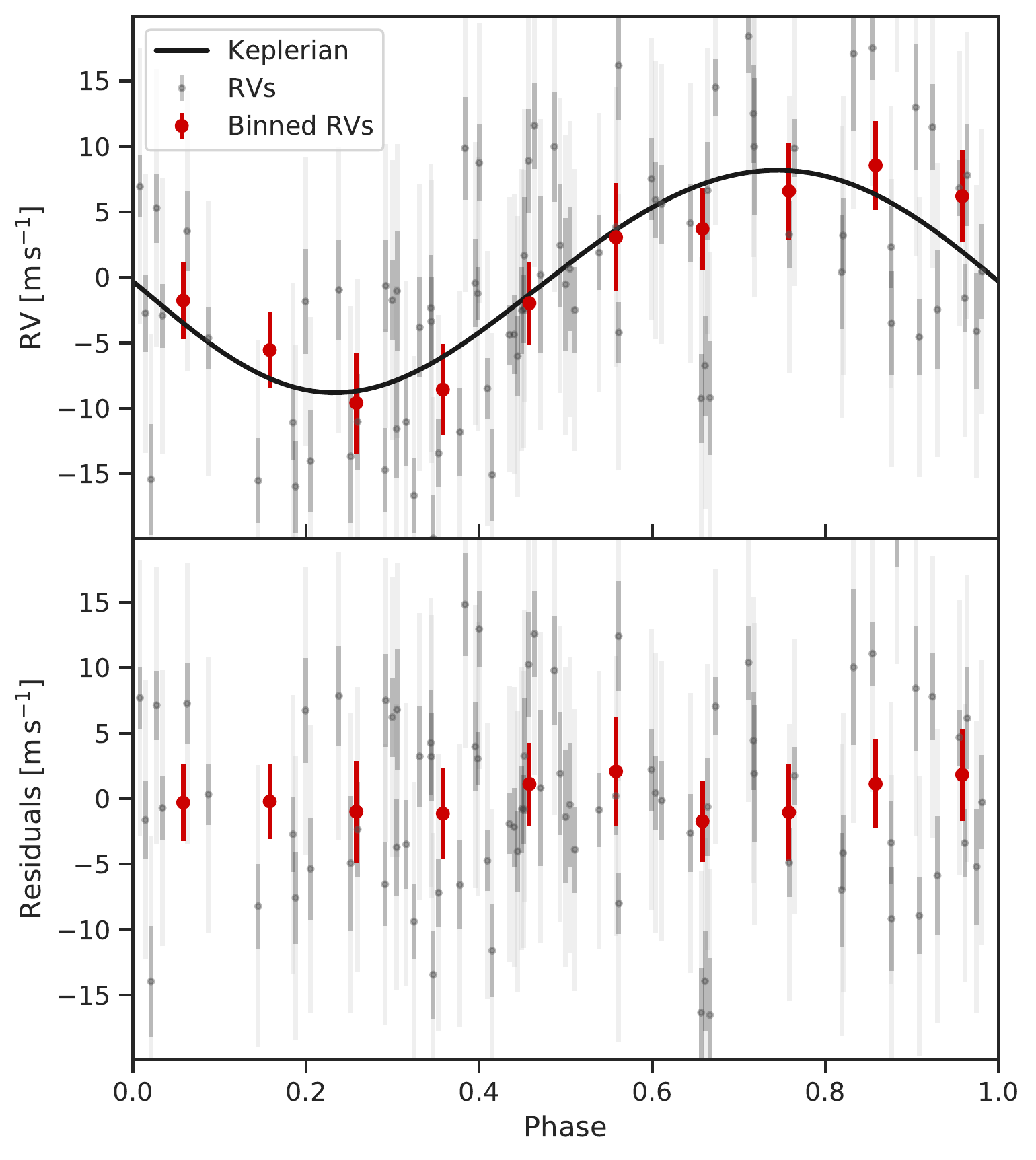}
    \caption{Phase-folded RV signals for \targetb\ (left) and \targetc\ (right) after subtracting the systemic velocity, the activity signal, and the other planet. 
    Grey circles with error bars show the HARPS measurements, and the error bars taking into account the jitter are shown with lighter grey. The black line show the best-fit Keplerian model. Red circles show binned RVs to facilitate comparison between data and models.
    }
    \label{fig:rvfolded}
\end{figure*}

We report fitted values and upper and lower uncertainties for each parameter in the final column of Table~\ref{tab:pars}, using the median and $16^{\rm th}$ and $84^{\rm th}$ percentiles of the marginal posterior distributions. The inferred model is shown alongside the data in Figure~\ref{fig:timeseries}. Figure~\ref{fig:rvfolded} shows the phase-folded individual RV signal for each planet. The signals of \targetb\ and \targetc\ are detected at the $2.3$ and $3.4\sigma$ level respectively, with amplitudes of \kb\ and \kc. While \targetc\ is detected at high confidence, the detection of \targetb\ is more tentative in this dataset. The eccentricities are consistent with zero.

The GP hyper-parameters are well constrained, and consistent with prior information about the rotation and activity behaviour of \target. The GP period is \pgp\, $=$ \jPGPtext\,d, consistent with the rotation period $P_{\rm rot}= 4.863\pm0.010$\,d reported by \citetalias{Plavchan2020}. The GP evolution time-scale \lbe\, $=$ \jlambdae\,d is 20 times the rotation period and is consistent with estimates from previous studies where a GP was used to describe the RVs only \citepalias{Plavchan2020,Klein2021,Cale2021}. The inverse harmonic complexity \lbp$\,=$ \jlambdap\ is relatively low. This can arise for a number of reasons: if the distribution of active regions on the stellar surface was relatively complex at the time of the observations, or if the latent variable $G(t)$ displays beat patterns caused by either differential rotation or the evolution of individual active regions (we note \citetalias{Klein2021} detected solar-like differential rotation using Doppler Imaging with SPIRou). 

We now examine the behaviour of the parameters $A_V$, $B_V$, and $A_{\rm \alpha}$, which control the relationship between the observables and the underlying latent variable $G(t)$. As shown in Table~\ref{tab:pars} and Figure~\ref{fig:corner_plot}, the posterior for $A_V$ is consistent with zero at the $1.5\sigma$ level, while those for $B_V$ and $A_{\rm \alpha}$ are significantly non-zero. In other words, the RV time-series behaves primarily like $\dot{G}(t)$ while the DLW time-series behaves primarily like $G(t)$. This is consistent with the prediction of \citet{Aigrain2012} and with what has been observed for other young, rapidly rotating, active stars including K2-100 \citep{Barragan2019} and HD\,73583 \citep[][]{Barragan2021b}.

The jitter terms, $\sigma_{\rm RV}$ and $\sigma_{\rm DLW}$, can be used as diagnostics of the effectiveness (or otherwise) of our model at explaining the full dataset. In both cases, the posteriors medians are approximately ten times smaller than the amplitude of the dominant activity term ($B_V$ and $A_{\rm \alpha}$, respectively), but roughly twice the nominal measurement uncertainties. This illustrates the fact that our activity model explains most of the RV and DLW variations, but is not perfect. 

A widely used test of the quality of a model fit is to compare the root-mean-square (RMS) before and after subtracting the best-fit model. Caution must be used in interpreting such a comparison, as a low residual RMS can result from over-fitting as readily as from a particularly good model. On the other hand, an anomalously high residual RMS would be a robust indication that the model cannot fully explain the observations. The initial RMS of the RV observations is 137\,\ms, compared to 7.5\,\ms\ after subtracting our best-fit, combined activity and planetary signal model. 
The DLW time-series has an initial RMS of $45 \times 10^3$\,\mmss\, which shrinks to $30 \times 10^3$\,\mmss\ after subtracting the best-fit model (which, in the case of DLW, includes activity only). Thus, the RMS of both time-series is reduced  by 94\% for the RVs and 33\% for the DLW, indicating that our model explains most of the variability seen in the RVs and a significant fraction of that seen in the DLWs. The residual RMS is nonetheless significant compared to the median formal uncertainty of each time-series ($3.6$\,\ms\ and $4.8$\,\mmss, respectively). This highlights that our model, while useful, is incomplete. The remaining variability might be due to stellar signals which our model cannot account for, or to instrumental systematics. Either way, it is absorbed by the jitter terms which, as we can see from Table~\ref{tab:pars}, have values similar to the residual RMS (as one would expect).

We also test if the planetary signals are recovered individually in the data. We ran three different setups: one including only planet b, one only planet c, and the last one including both signals. The results of this test are summarised in Appendix~\ref{sec:planetarysignals}.
We found that we are able to recover each planetary signal individually, even if the signal of the other planet is not included in the model.
In addition, we found that the model producing the highest likelihood is the one including both planetary signals. However, we note, that if we use standard model comparison parameters such as Akaike Information Criterion and Bayesian Information Criterion, the preferred model is the one including \targetc\ signal only. This is expected given that the detection of \targetb\ is not well constrained. Nonetheless, we adopt the two planet model given that this is physically motivated. We know a priori that both planets are present from the transit observations.

\begin{table*}
\begin{center}
  \caption{Priors and posterior median and confidence intervals for the parameters of the joint RV \& DLW model. The posterior distributions from which we infer these parameters are shown in Figure~\ref{fig:corner_plot}. \label{tab:pars}}  
  \begin{tabular}{lcc}
  \hline
  \hline
  \noalign{\smallskip}
  Parameter & Prior$^{(\mathrm{a})}$ & Posterior value$^{(\mathrm{b})}$ \\
  \noalign{\smallskip}
  \hline
  \noalign{\smallskip}
  \multicolumn{3}{l}{\emph{\bf \targetb's parameters }} \\
  \noalign{\smallskip}
    Orbital period $P_{\mathrm{orb}}$ (days)  & $\mathcal{N}[8.463000,0.000002]$ &\Pb[] \\
    Transit epoch $T_0$ (BJD$_\mathrm{TDB}-$2\,450\,000)  & $\mathcal{N}[8330.39051 , 0.00015]$ & \Tzerob[]  \\  
    Eccentricity $e$  & $\mathcal{B}[1.52,29]^{(\mathrm{c})}$ & \eb[]  \\
    Angle of periastron $\omega$ (deg) &  $\mathcal{U}[0,360]$ & \wb[]  \\
    Doppler semi-amplitude variation $K$ (m s$^{-1}$) & $\mathcal{U}[0,50]$ & \kb[] \\
    \multicolumn{3}{l}{\emph{ \bf \targetc's parameters}} \\
    Orbital period $P_{\mathrm{orb}}$ (days)  & $\mathcal{N}[18.859019,0.000016]$ &\Pc[] \\
    Transit epoch $T_0$ (BJD$_\mathrm{TDB}-$2\,450\,000)  & $\mathcal{N}[8342.2223,0.0005]$ & \Tzeroc[]  \\ 
    Eccentricity $e$  & $\mathcal{B}[1.52,29]^{(\mathrm{c})}$ & \ec[]  \\
    Angle of periastron $\omega$ (deg) &  $\mathcal{U}[0,360]$ & \wc[]  \\
    Doppler semi-amplitude variation $K$ (m s$^{-1}$) & $\mathcal{U}[0,50]$ & \kc[] \\
    \multicolumn{3}{l}{\emph{ \bf Multidimensional GP parameters}} \\
    GP Period $P_{\rm GP}$ (days) &  $\mathcal{U}[4.5,5.5]$ & \jPGP[] \\
    $\lambda_{\rm p}$ &  $\mathcal{U}[0.1,2]$ &  \jlambdap[] \\
    $\lambda_{\rm e}$ (days) &  $\mathcal{U}[10,500]$ &  \jlambdae[] \\
    $A_{\rm V}$ (\ms)  &  $\mathcal{U}[0,100]$ & \jAzero \\
    $B_{\rm V}$ (\ms\,d) &  $\mathcal{U}[-1000,1000]$ & \jAone \\
    $A_{\alpha}$ (1000 \mmss) &  $\mathcal{U}[0,1000]$ & \jAtwo \\
    \multicolumn{3}{l}{\emph{ \bf Other parameters}} \\
    $C_v$ (Offset  RV, \kms) & $\mathcal{U}[ -0.82 , 0.71]$ & \RVs[] \\
    $C_\alpha$ (Offset DLW, \mmss) & $\mathcal{U}[-1,1]$ & \DLW[]  \\
    Jitter term $\sigma_{j,\rm RV}$ (\ms) & $\mathcal{J}[0,1000]$ & \jRVs[] \\
    Jitter term $\sigma_{j,\rm DLW}$  (1000 \mmss)  & $\mathcal{J}[0,1000]$ & \jDLW[] \\
    \noalign{\smallskip}
    \hline
    \multicolumn{3}{l}{\emph{\bf Derived Parameters}} \\
    \multicolumn{3}{l}{\emph{\bf \targetb's derived Parameters}} \\
    Planet mass $M_\mathrm{p}$ ($M_{\rm \oplus}$) & $\cdots$ & \mpb[]  \\
    Planet density $\rho_\mathrm{p}$ (${\rm g\,cm^{-3}}$) & $\cdots$ & \pdenb[]  \\
    \multicolumn{2}{l}{\emph{\bf \targetc's derived Parameters}} \\
    Planet mass $M_\mathrm{p}$ ($M_{\rm \oplus}$) & $\cdots$ &  \mpc[]  \\
    Planet density $\rho_\mathrm{p}$ (${\rm g\,cm^{-3}}$) & $\cdots$ & \pdenc[]  \\
    \hline
  \end{tabular}
\end{center}
  \begin{tablenotes}
  \item \emph{Note} -- $^{(\mathrm{a})}$ $\mathcal{U}[a,b]$ refers to uniform priors between $a$ and $b$, $\mathcal{N}[\mu,\sigma]$ to Gaussian priors with mean $\mu$ and standard deviation $\sigma$, $\mathcal{B}[a,b]$ to a beta distribution with shape parameters $a$ and $b$,  and $\mathcal{J}[a,b]$ is the modified Jeffrey's prior as defined by \citet[eq.~16]{Gregory2005}.
  $^{(\mathrm{b})}$  Inferred parameters and errors are defined as the median and 68.3\% credible interval of the posterior distribution. 
   $^{(\mathrm{c})}$  Beta distribution to inform eccentricity sampling using the beta distribution for multi planetary systems as defined by \citet[][]{VanEylen2019}.
\end{tablenotes}
\end{table*}

\section{Discussion}
\label{sec:discuss}

In this Section, we discuss the implications of the results presented in Section~\ref{sec:res} for the masses, orbits, composition and evolution of \targetb\ and c.

\subsection{Planet masses}

From our estimates for $K_{\rm b}=$ \kb\ and $K_{\rm c}=$ \kc, we directly obtain mass estimates for the two planets of $M_\mathrm{b}=$ \mpb\ and $M_\mathrm{c}=$ \mpc. This represents a $3.4\sigma$ detection of planet c, but only a tentative, $2.3\sigma$ detection of planet b.

\citetalias{Klein2021} reported a mass estimate of $17.1\pm4.6\,M_\oplus$ for planet b, which is consistent at the $1\sigma$ level with our estimate. We note that \citetalias{Klein2021} did not include \targetc\ in their analysis, since the second planet was not confirmed at the time. 

More recently, \citetalias{Cale2021} reported a mass of $20.1\pm1.6\,M_\oplus$ for planet b and a $5\sigma$ upper limit for planet c of $M_{\rm c}<20.1\,M_\oplus$, using a combination of data from multiple optical and near-IR spectrographs, and a model which includes only the RVs, but exploits the chromaticity of the activity signal. Our results are not consistent with these values. Although it is difficult to identify the exact source of the discrepancy, several  factors are likely at play. The first is the difference in the wavelength coverage; including both optical and near-IR data should in principle give a better handle on the activity signals, though it also necessitates an increase in the number of free parameters in the model. Another difference is the time-sampling of the datasets used, which is much sparser in the case of \citetalias{Cale2021} than in the present work. Sparser time sampling severely limits the ability to model the activity signal. The third is the fact that we model an activity indicator simultaneously with the RVs, which again should help constrain the activity component of the model. We note that \citetalias{Cale2021} performed two different analyses, with different assumptions about the chromaticity of the activity signal, and that these gave mutually inconsistent results, particularly for the mass of planet c. 

Another difference is that \citetalias{Cale2021} use a Gaussian prior with mean $0.19$ and standard deviation $0.04$ for the eccentricity of planet b, derived from \textit{Spitzer} secondary eclipse observations. As the latter are not yet published, we did not incorporate this prior in the analysis presented in Section~\ref{sec:res}, but we did carry out an additional model run with the same eccentricity prior for planet b as used by \citetalias{Cale2021}. This did not appreciably affect our mass estimates for either planet.

Finally, exploring the parameter space for such complex, activity plus Keplerian models, is in general challenging. To check how robust our results are to the details of the model, we carried out simple tests including only planet b or only planet c, and found that the resulting RV semi-amplitudes were consistent with the values obtained from the model including both planets. We also carried out a number of additional tests using simulations, described in Section~\ref{sec:sims}, to further establish confidence in our results. Overall, however, it is clear that  continued RV monitoring of the system, as well as independent analyses of all the available datasets by different teams, would be desirable to resolve the discrepancies between the  published mass estimates.

\subsection{TTV analysis}

As \targetb\ and c are close to a 2:1 mean-motion resonance, transit timing variations (TTVs) can be used to constrain their masses and eccentricities.  \citet{Szabo2021} and \citet{Gilbert2021} report tentative TTVs for \targetb\ at the level of 3--4\,min and 80\,s, respectively, using \tess\ and \cheops\ observations. Although the measurement of transit times is somewhat difficult because of \target's frequent flaring, these results indicate that strong TTVs, with amplitudes in excess of $\sim 10$\,min, are not present in the system. This can be used to provide a sanity check on the mass and eccentricity estimates derived in this work and in previous RV analyses, and to motivate future TTV follow-up. 

Using the TTVFast package \citep{Deck2014} we forward modelled the TTVs expected from a two-planet configuration as observed in the AU Mic system. TTV amplitudes are most sensitive to planetary mass and eccentricity, so we varied those on a grid over a plausible range for these parameters, noting the amplitude of the TTV signal in each case. Our predictions are summarised in Appendix~\ref{sec:TTV_pred}. Even for the most extreme mass ratio and eccentricities compatible with the results of our RV analysis, the maximum amplitude of the TTVs never exceeds a few minutes, with a super-period of ~80 days. We therefore conclude that our mass estimates are consistent with existing transit observations.

\subsection{Dynamical analysis}

We used \texttt{mercury6} \citep{Chambers1999} to test the dynamical stability of the \target\ system. We used the masses and orbital parameters reported in Table~\ref{tab:pars}, assuming that both planets have co-planar orbits. Starting from this configuration, we evolved the system for 1\,Gyr with steps of 0.5\,d per integration. We found that the eccentricity and semi-major axis of both planets display  periodic fluctuations with a period of $\sim 1000$ years, but these fluctuations are contained and the system appears stable over the full duration of the simulation. The eccentricity of \targetb\ oscillates between 0.05 and 0.25, and its semi-major axis varies by $8\times10^{-5}$\,AU. The eccentricity of \targetc\ remains $< 0.10$ and the maximum departure of its semi-major axis from the starting value is $1.3\times10^{-4}$\,AU. 
We conclude that the orbital parameters and masses derived for \target\ planets are consistent with a dynamically stable system.

\subsection{Tests to establish detection robustness}
\label{sec:sims}

\begin{figure*}
    \centering
    \includegraphics[width=0.49\textwidth]{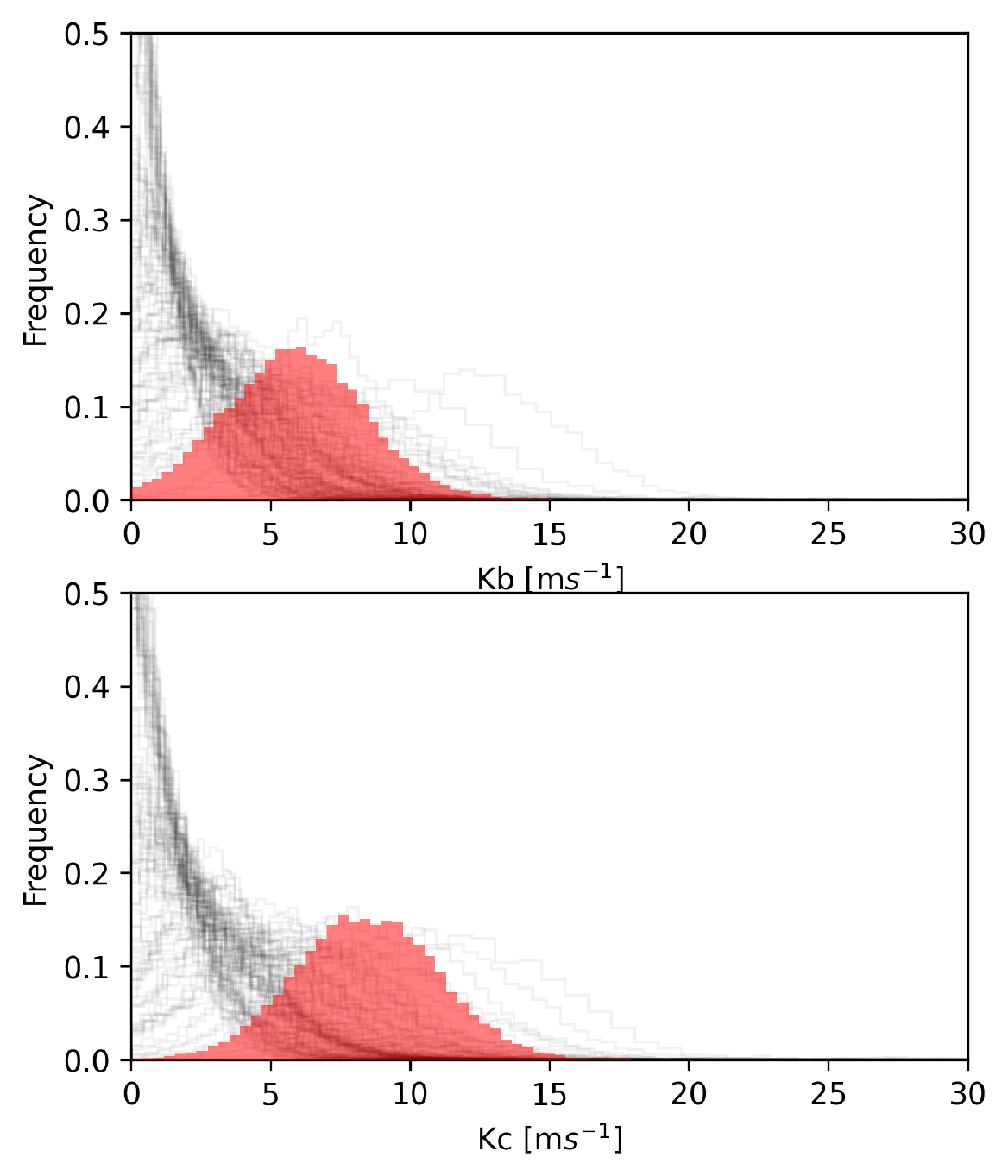}
    \includegraphics[width=0.49\textwidth]{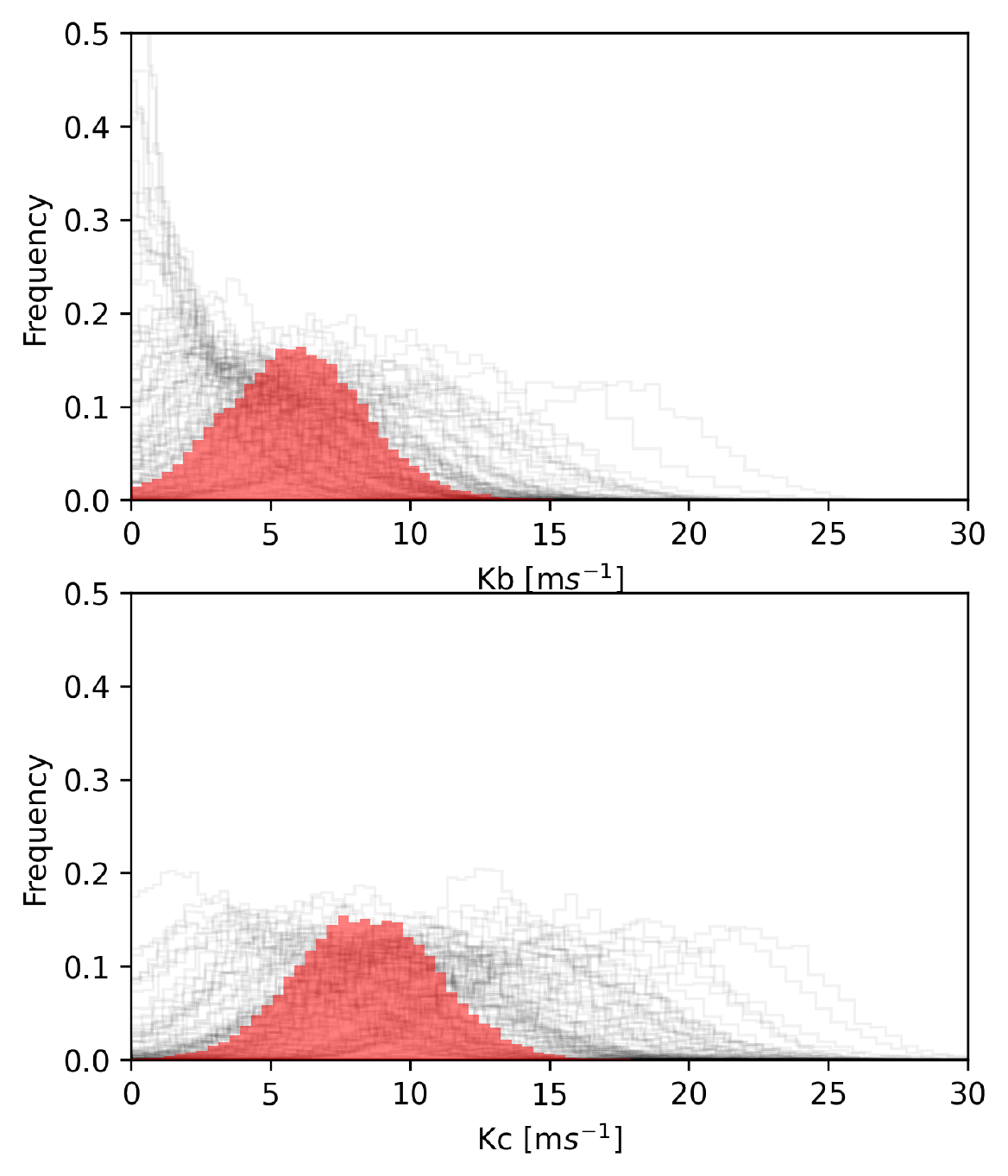}
    \caption{Results of the activity-only (left) and activity plus Keplerian (right) simulations (see Section~\protect\ref{sec:sims} for details). In each panel, the red histogram shows the posterior derived from the original data, and the thin grey lines show the posteriors obtained from the simulations.}
    \label{fig:recovery}
\end{figure*}

It has been shown that the combination of complex models and the window function of the observations can create spurious planet-like signals in RV time-series, specially for active stars \citep[e.g.,][]{Rajpaul2016}. In order to check the reliability of our RV detection, we performed numerical simulations similar to \citet{Barragan2019}. We used \citlalatonac\ \citepalias{pyaneti2} to simulate an RV and DLW time-series containing only the best-fit activity model as obtained from the real data. 
We decided to use the median values of our derived parameters given that the posteriors are quasi-Gaussian uni-modal distributions. We then added correlated noise using a squared exponential kernel with a length-scale of one day, and the same amplitude as the jitter term obtained from the real data, plus white noise for each observation according to its nominal measurement uncertainty. We did this 100 times to obtain 100 simulated activity-only time-series, with similar noise properties and the same time-sampling as the real data.

We modelled each activity-only synthetic dataset using a two-planet and 2-dimensional GP configuration as described in Sect.~\ref{sec:modellingmcmc}. For each simulation, we plot the posterior over the semi-amplitudes for both `planets' in the left column of Figure~\ref{fig:recovery}, compared to the posterior obtained from the real dataset. We then count the fraction of the simulations where the recovered semi-amplitude for planets b and c equals or exceeds the median value obtained from the actual observations. 
For planet b, this occurred 11\% if the time in the 100 simulations. For planet c, it occurred in 2\% of the cases. Taking Poisson counting errors into account, this is fully consistent with the confidence intervals that we derived for the real observations.

We then repeat the same exercise, creating another 100 synthetic datasets, but this time, injecting two Keplerian signals with the median planet parameters obtained from the real dataset, as reported in Table~\ref{tab:pars}. Again, we modelled these synthetic datasets using \pyaneti\ with the same configuration as described in Sect.~\ref{sec:modellingmcmc}. The resulting posteriors are shown in the right column of Figure~\ref{fig:recovery}. In this case, we are interested in two distinct questions. First, how often are \targetb\ and \targetc\ detected at a given confidence level from these simulations? For this purpose, we consider that a detection has occurred if the median of the posterior is larger than $2\sigma$, where $\sigma$ is half the interval between the $16.5^{\rm th}$ and $83.5^{\rm th}$ percentiles. We find that \targetb\ is detected 42\% of the time, and \targetc\ 88\% of the time. This confirms that, if the planet masses are similar to the maximum a posteriori values we derived from the real data, more data is needed for a robust detection of planet b.
Second, we asked what fraction of the time the recovered semi-amplitude is within $2\sigma$ of the injected value? We find that this happens 91\% and 89\% of the time, for \targetb\ and \targetc\ respectively. For Gaussian posteriors, we would expect these numbers to be around 95\%. Again, accounting for Poisson counting errors, this result is thus consistent with what we expect.

\subsection{Composition and evolution}
\label{sec:evolution}

\begin{figure*}
    \centering
    \includegraphics[width=0.95\textwidth]{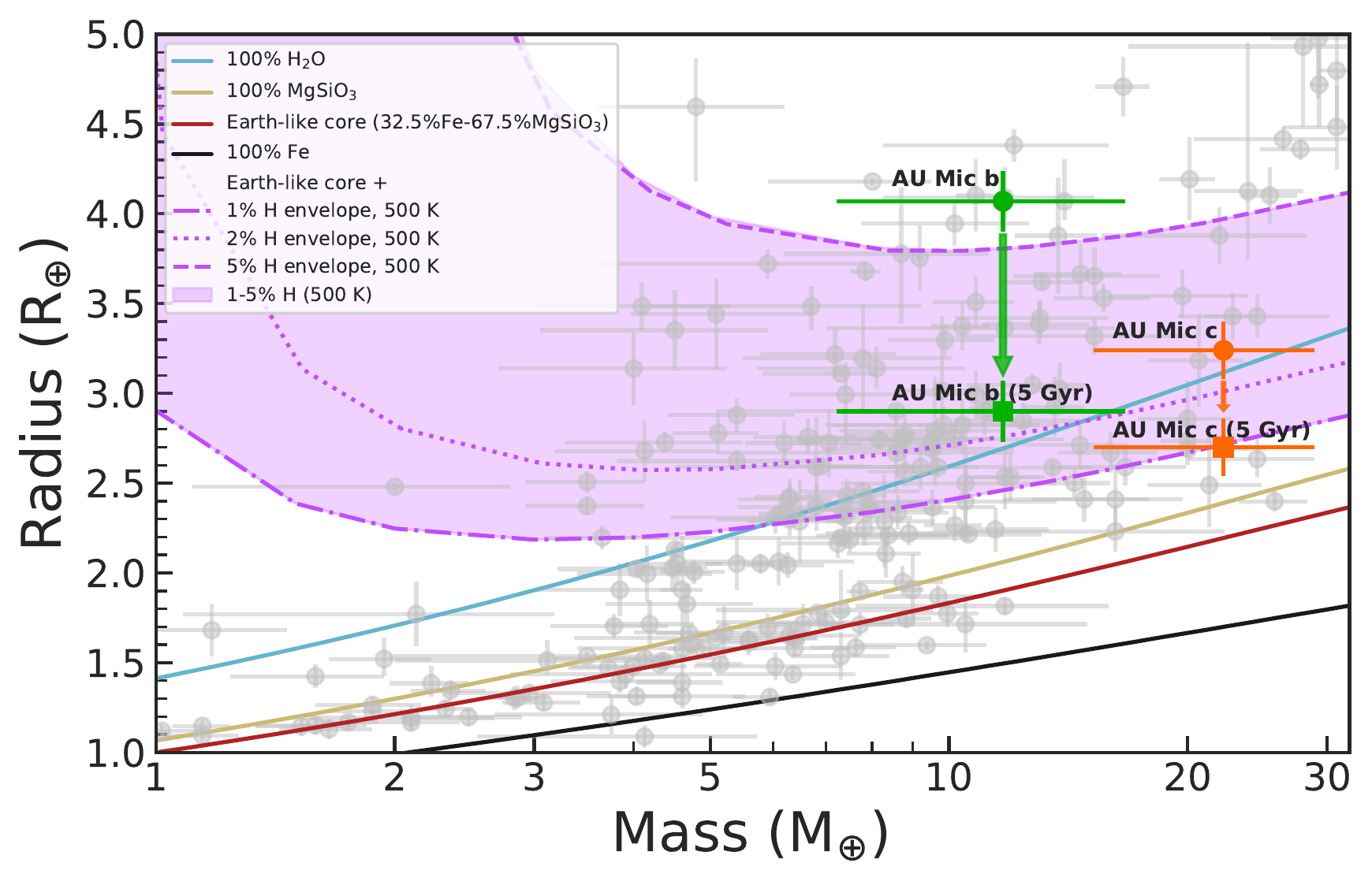}
    \caption{Mass \emph{vs} radius diagram for small exoplanets ($1 < R_{\rm p} < 5 R_\oplus$ and $1 < M_{\rm p} < 32 M_\oplus$). Grey points with error bars show planets with mass and radius measurements better than 50\% \citep[As in the TEPCAT catalogue, \url{https://www.astro.keele.ac.uk/jkt/tepcat/}, ][]{Sotuhworth2007}.
    \targetb\ and c are shown with a green and orange circle, respectively. The green and orange arrows indicate the expected evolutionary pathway for the planets and their final state after 5\,Gyr.
    The green and orange squares show the expected position of AU Mic b and c after 5 Gyr, respectively.
    Solid lines represent two-layer models as given by \citet{Zeng2016} with a different colour corresponding to a different mixture of elements. Purple non-solid lines correspond to Earth-like rocky cores surrounded by a Hydrogen envelope of 1\% (dash-dotted line), 2\% (dotted line), and 5\% (dashed) Hydrogen mass for {\it evolved} exoplanets assuming an equilibrium temperature of 500\,K \citep{Zeng2019}.
    \textbf{Note}: One should compare the expected evolved position of the planets with the composition curves, rather than their current position.
    }
    \label{fig:mr}
\end{figure*}

\begin{figure*}
    \centering
    \includegraphics[width=0.49\textwidth]{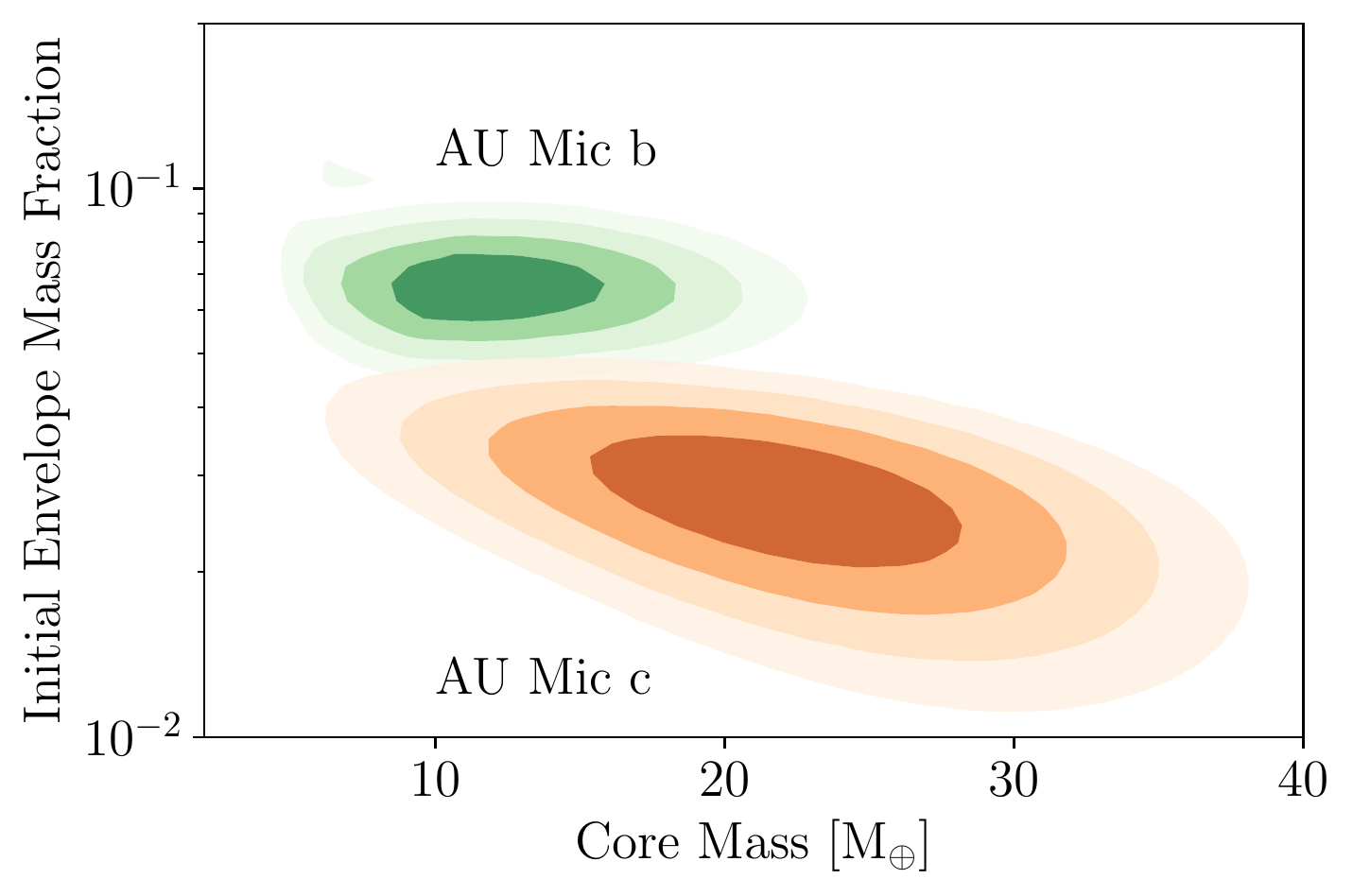}
    \includegraphics[width=0.49\textwidth]{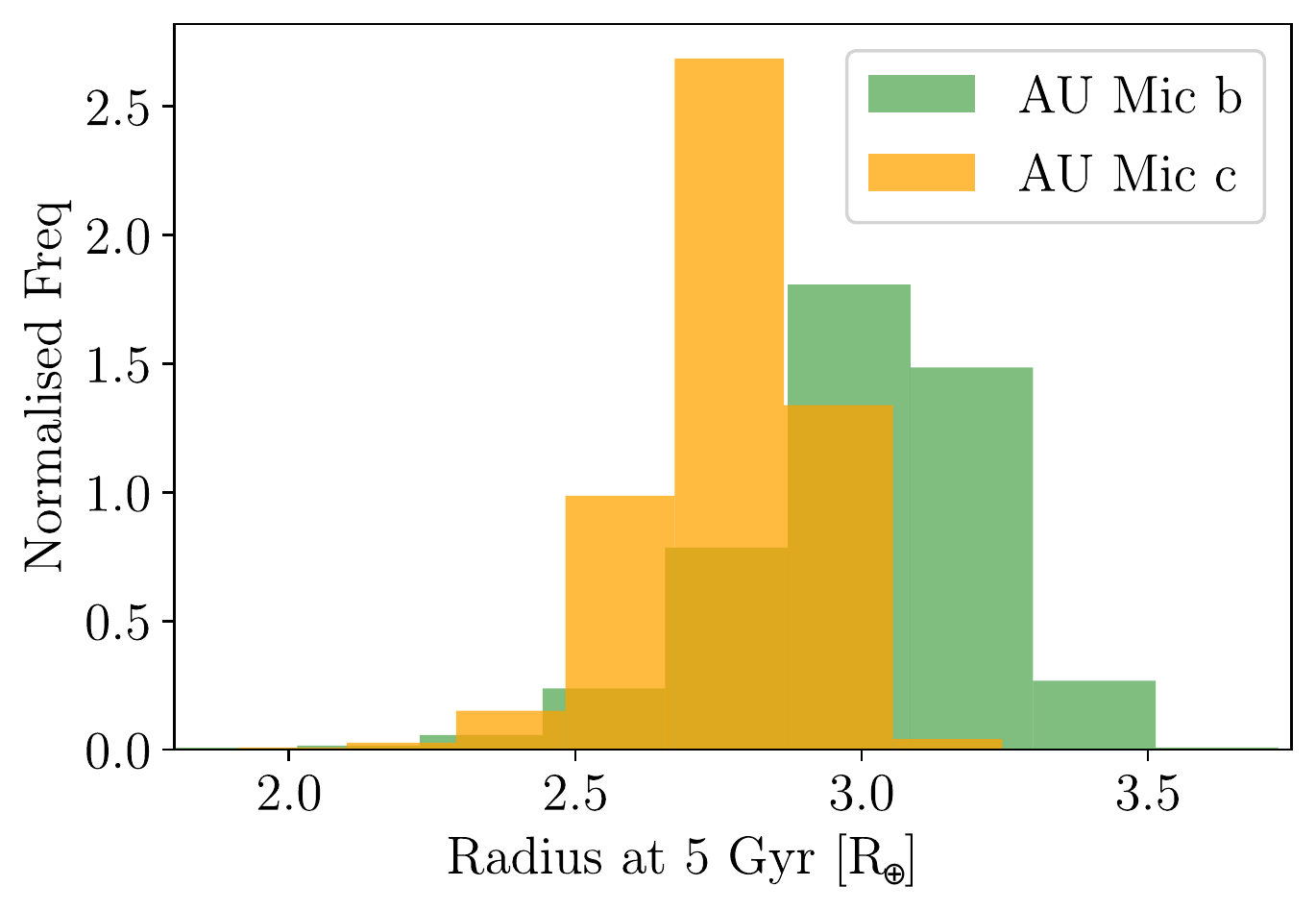}
    \caption{Initial envelope mass fractions and core masses (left - with contours shown at $0.5\sigma$ intervals) and radii after 5\,Gyr (right) for \targetb\ and c (in green and orange, respectively) compatible with the measured masses and radii, based on evolutionary calculations including cooling and contraction and photo-evaporation (see Section~\protect\ref{sec:evolution} for details).}
    \label{fig:evolution}
\end{figure*}

Having measured radii and masses for both planets allows to place them in a mass-radius (M-R) diagram, as shown in Figure~\ref{fig:mr}. Also shown on that figure are other planets with well-determined masses and radii and theoretical mass-radius relations for  terrestrial and ocean  worlds \citep{Zeng2016}, and for {\it evolved} planets with H/He envelopes \citep{Zeng2019} accounting for 1--5\% of the total mass, at a temperature of 500\,K, which is close to the equilibrium temperatures of both \targetb\ and c (\eqtempb\ and \eqtempc\ respectively, see \citealt{Martioli2020} and Table~\ref{tab:literature}). We note that although it is not ideal to directly compare these young planets with their older counterparts, this is standard practice in the field.


Comparison to these theoretical composition curves for evolved, $\sim$Gyr old planets indicates that planet c is compatible with a pure H$_2$O world, and planet b requires an atmosphere containing some H/He. However, it is difficult to imagine how a pure H$_2$O planet could form. Coupled with the fact planet b hosts an atmosphere containing H/He, we consider it more likely that both planets possess voluminous H/He atmospheres. However, comparison to these evolved mass-radius relationships is indicative only. Young planets still contain considerable thermal energy left over from their assembly \citep[e.g.][]{Ginzburg2016}, resulting in significantly larger planetary radii compared to evolved planets with the same composition \citep[e.g.][]{Owen2020}.

Thus, we compare the measured masses and radii to evolutionary models for close-in exoplanets, that include the impact of cooling and contraction of any H/He atmosphere and photoevaporative loss using {\sc mesa} models \citep{Owen2013}. The evolutionary calculations are identical to those used in \citet{Owen2020} and \citet{Mann2021}. In this manner, the measured masses and radii of planets b and c, can be used to constrain their possible evolutionary pathways, and hence their composition and thermodynamic state both today and at  their formation.

We find that both planets formed with voluminous, but low mass H/He dominated atmospheres ($\sim 8\%$ and $\sim 3\%$ by mass for planets b and c, respectively, as shown in the left panel of Figure~\ref{fig:evolution}), and as such are likely progenitors of the ubiquitous super-Earth/sub-Neptune population. This is further evidenced when we evolve these planets forward to an age of 5\,Gyr finding that in the vast majority of cases these planets retain a fraction of their H/He atmosphere, reaching a radius of $\sim$ 2.9 and $\sim 2.7\,R_\oplus$ for b and c respectively (see Figure~\ref{fig:evolution}, right panel). However, there are two standout results. The first is that the amount of H/He these planets accreted from their parent discs is significantly less than predicted from standard models of core-accretion \citep[e.g.][]{Lee2015}. Based on those models, we would have expected planet b, for example, to have accreted between 25 and 75\% of its mass in H/He. This fraction is high enough that it could plausibly have undergone runaway accretion. The envelope mass fraction we infer from the measured mass and radius of planet b is 3--10 times smaller than this theoretical prediction. The second standout result is the trend between the two planets, where planet c has a higher inferred core-mass, but accreted less H/He than planet b. This is exactly the opposite of what is expected from core-accretion, where more massive cores accrete more H/He \citep[e.g.][]{Pollack1996}. 

There have already been indications that close-in planets may accrete less H/He than predicted by core-accretion, based on a sample of old individual planets \citep{Jankovic2019}, and from inferences about the population of planets discovered by {\it Kepler} \citep{Rogers2021}. However, this is the first time it has been directly identified in young planets that have not been significantly affected by atmospheric mass-loss. Possible mechanisms to resolve this tension with core-accretion include forming the planets late in the disc's lifetime, during the dispersal phase \citep[e.g.][]{Ikoma2012,Lee2016}, or the conclusion from 3D simulations that high-entropy gas is continually recycled into the forming atmosphere, preventing it from cooling and accreting more gas \citep[e.g.][]{Ormel2015,Chen2020b,AliDib2020}. Finally, additional mass-loss during disc dispersal can dramatically remove accreted material, as the disc's pressure confinement of the proto-atmosphere is rapidly removed \citep[e.g.][]{Owen2016,Ginzburg2016}.

However, none of these proposed mechanisms can reconcile core-accretion with the observation that planet c, with its more massive core, started off with a smaller H/He atmosphere. This is even more curious as planet c appears massive enough to have undergone run-away accretion, even at its current short-period orbit, and formed a giant planet \citep[e.g.][]{Rafikov2006,Lee2014}. However, it only had an initial H/He mass fraction of a few percent.

One possible explanation is that the two planets formed further from the star than their present-day locations, then migrated inwards. Standard core accretion models predict that the mass of the accreted atmosphere grows with core mass as $\sim M_{\rm core}^{1.7}$ \citep[as evidenced by][]{Lee2015}, and is approximately logarithmically sensitive to the disc pressure \citep[e.g.][]{Piso2014}. Applying these scaling laws directly to the core and envelope masses we infer for \targetb\ and c, it would imply that the two planets formed in regions of the disc with extremely different densities (by a factor $\gtrsim \exp(10)$). Such a scenario would also require that planet c did not start migrating inwards until the disc dispersed (otherwise it would have accreted more gas). It is then difficult to see how the planet would have reached its current orbit by the present age of the system, in the absence of a gas disc. On the other hand, the critical core mass for runaway gas accretion depends on the rate at which the core is accreting solid material, which itself depends on location within the disc \citep{Terquem2014}. This opens up the possibility of a scenario where the planets form at different locations in the disc, but less extremely so, and migrate inwards during the late stages of the disc's dispersal. Exploring this scenario further using detailed calculations is, however, beyond the scope of the present work.

Another possible explanation would be that planet c started forming later, and thus acquired its envelope over a shorter time, than planet b. However, the accreted mass has a very weak dependence on accretion time ($M_{\rm env} \propto t^{0.4}$, \citealt{Lee2015}). Following the same logic as the gas surface density arguments given above, this would require planet c to form over a time period $10^{1/0.4} \sim 300$ times shorter than planet b. Again, this would require a great deal of fine tuning.

A second, speculative scenario that would reconcile both c's large core-mass and lower H/He atmosphere mass compared to b, is that c is the product of a post-disc-dispersal giant impact. Giant impacts are not an unlikely outcome for planets that find themselves on short period orbits after disc-dispersal \citep[e.g.][]{Poon2020,Bonomo2019}. Giant impacts can remove large amounts of H/He in any collision \citep[e.g.][]{Liu2015,Inamdar2016} in addition to the fact that the collision inflates any residual H/He atmosphere resulting in enhanced photoevaporation \citep{Biersteker2019}. Planets in multiple systems found by {\it Kepler} tend to have similar radii \citep[e.g.][]{Weiss2018} and masses \citep[e.g.][]{Millholland2017} to each other. This has been interpreted as evidence for uniformity in the underlying core-mass within each system \citep{Millholland2021}. Thus, if \target\ formed three planets with properties similar to b, and two underwent a giant impact, a planet with properties similar to c would be the natural outcome. One way to test this speculative scenario is to measure the obliquity of planet c. Any indication of difference between the spin-orbit angles of the two planets \citep[the orbital plane of planet b having been found aligned with the stellar rotation axis;][]{Addison2021, Martioli2020, Szabo2021, Hirano2020} would indicate that significant dynamical interaction between planets in the system took place after disc-dispersal. The amplitude of the Rossiter-McLaughlin (RM) effect induced by planet c is expected to be of order $\sim 10$\,\ms, which is feasible with current high-precision spectrographs.

\section{Conclusions}
\label{sec:concl}

We have analysed data from an intensive monitoring campaign using HARPS with multidimensional GP framework \citepalias{pyaneti2}, and measured the masses of \targetb\ and c to be \mpb\ and \mpc. We detect planet c at $3.4\sigma$ confidence and planet b at $2.3\sigma$ confidence. Our results are insensitive to the choice of RV time-series (DRS, \serval, \terra) and to the specific activity indicators used, though this choice affects the degree of confidence in the detections slightly. Our mass measurements, combined with literature radius estimates, indicate that \targetb\ has a significant H/He envelope and may have an internal heat source. On the other hand, we find that \targetc\ is denser and compatible with either a rocky core surrounded by a $\sim 2\%$ H/He envelope or a pure H$_2$O composition. These results are in tension with current core-accretion models, which predict that planet c, with a more massive core, should have the larger H/He envelope of the two. We provide a possible, speculative explanation for this discrepancy involving a giant impact, but stress that additional observations are needed to further refine the mass measurements and constrain the obliquity of planet c.

The RV signals of the two planets around \target\ are around 50 times smaller than the activity signals. Disentangling the former from the latter is very challenging, as highlighted by the discrepancies between our results and those of other teams working on the same system using different instruments and methods, which need to be better understood. Further observations of this system and independent analyses of the available datasets are needed to resolve these discrepancies. However, the fact that these detections were possible at all is very encouraging for the continued characterisation of the increasing number of young transiting planets being discovered by surveys such as \ktwo\ and \tess. The characterisation of planets around young stars also provides an extreme test case for activity mitigation techniques in RVs. The methods being developed in this context will prove useful for future searches for smaller and cooler planets around less active stars, where the amplitude ratios between planetary and activity signals are similar.

\section*{Acknowledgements}

We gratefully acknowledge X.\ Dumusque and F.\ Bouchy for their coordination of the HARPS time-share, which allowed us to spread the observations over the two semesters with the appropriate time-sampling, and all the observers involved in the time-share (F.\ Bouchy, B.\ Canto, I.\ de Castro, G.\ Dransfield, M.\ Esposito, V.\ Van Eylen, F.\ Hawthorn, M.\ Hobson, V.\ Hodzic, D.\ Martin, J.\ McCormac, H.\ Osborne, J.\ Otegi, A.\ Suarez, P.\ Torres) for carrying out the observations on our behalf.
We also thank J.\ Patterson for managing the Oxford Astrophysics compute cluster, \textit{glamdring}, which was used to carry out the data analysis and simulations. In addition, we thank Caroline Terquem for her insights and suggestions regarding the formation of the planets.
This study is based on observations collected at the European Southern Observatory under ESO programmes 0105.C-0288\footnote{The P105 observations were delayed because of the COVID-19 pandemic and taken in P107.} \& 0106.C-0852 (PIs Aigrain and Zicher).
This research has made use of NASA’s Astrophysics Data System.
NZ acknowledges support from the UK Science and Technology Facilities Council (STFC) under Grant Code ST/N504233/1, studentship no. 1947725.  
This publication is part of a project that has received funding from the European Research Council (ERC) under the European Union’s Horizon 2020 research and innovation programme (Grant agreement No. 865624). 
L.D.N thanks the Swiss National Science Foundation for support under Early Postdoc.Mobility grant P2GEP2\_200044.
JEO is supported by a Royal Society University Research Fellowship. This project has received funding from the European Research Council (ERC) under the European Union’s Horizon 2020 research and innovation programme (Grant agreement No. 853022, PEVAP). 
Some of this work was performed using the DiRAC Data Intensive service at Leicester, operated by the University of Leicester IT Services, which forms part of the STFC DiRAC HPC Facility (www.dirac.ac.uk). The equipment was funded by BEIS capital funding via STFC capital grants ST/K000373/1 and ST/R002363/1 and STFC DiRAC Operations grant ST/R001014/1. DiRAC is part of the National e-Infrastructure.
{NZ, OB, and BK would like to thank the pizzeria and caf\'e in Oxford that gave us the energy to detect the two planetary signals during COVID-19 times. }

\section*{Data Availability}

The HARPS spectra used in this study are available on the ESO archive\footnote{\url{http://archive.eso.org/eso/eso_archive_main.html}}. The radial velocities and activity indicators extracted with the DRS, \terra\ and \serval\ are listed in Table~\ref{tab:rvs}, a machine-readable version of which is provided in the Supplementary Material.
The code used to perform the evolutionary analysis is available at \mesaplanet \footnote{\url{https://github.com/jo276/MESAplanet}\faGithub}.
Pyaneti is available at \pyaneti
\footnote{\url{https://github.com/oscaribv/pyaneti}\faGithub}.



\bibliographystyle{mnras}
\bibliography{bibs} 




\appendix



\section{Correlations plot}

\begin{figure*}
    \centering
    \includegraphics[width=\textwidth]{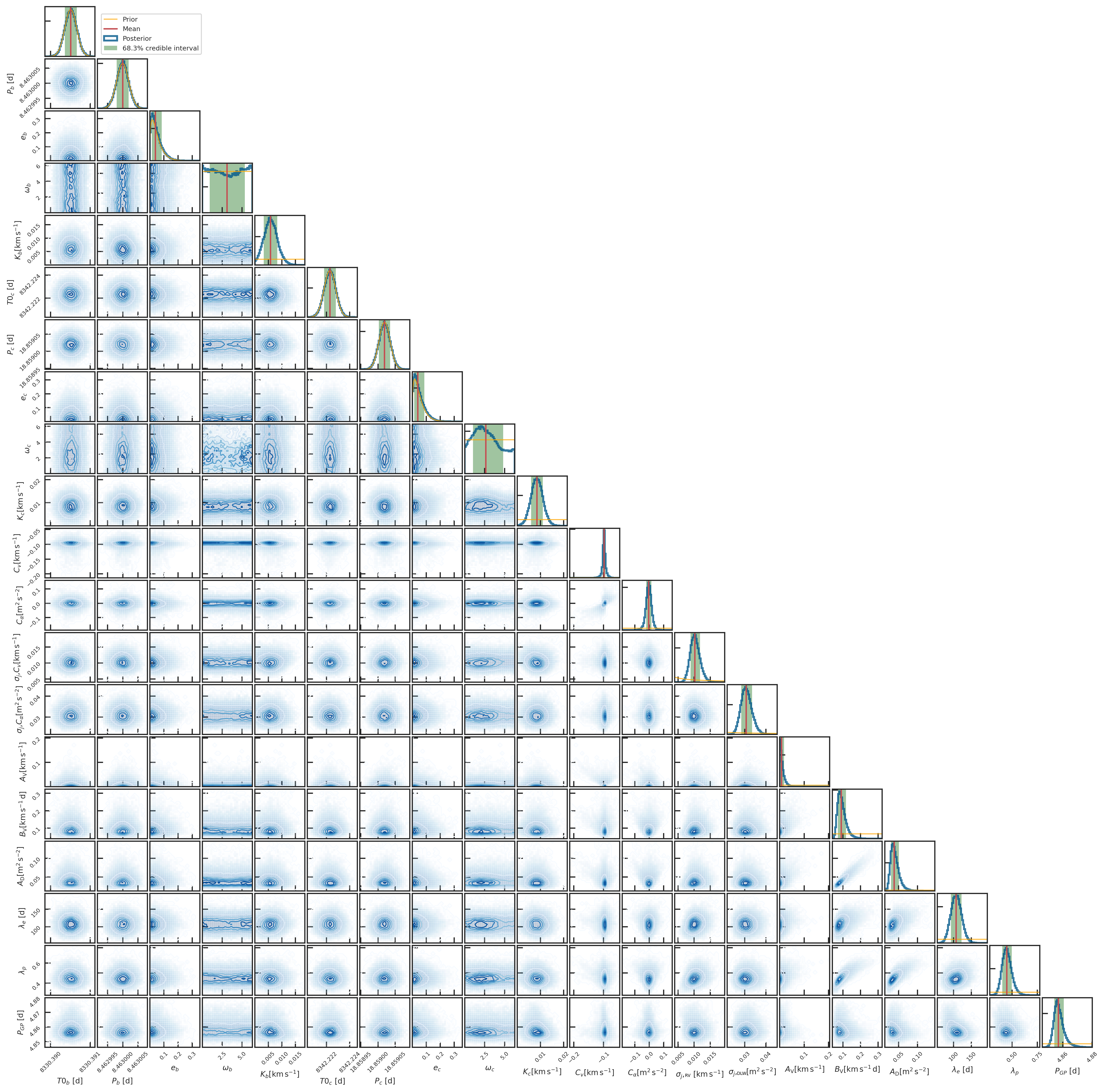}
    \caption{Posterior and correlation plots for the sampled parameters (MCMC corner plot). The inferred parameters from these posteriors are given in Table~\ref{tab:pars}.}
    \label{fig:corner_plot}
\end{figure*}

Figure~\ref{fig:corner_plot} shows the posterior and correlations plots for the sampled parameters of the described in Sect.~\ref{sec:modellingrvs}. The parameters correspond to the SERVAL RVs and DLW model (Model 3 in Sect.~\ref{sec:choice_act}).

\section{Activity models}
\label{sec:threeruns}

In Table~\ref{tab:parsmodels} we show the priors and inferred parameters of the three models described in Section~\ref{sec:choice_act}.

\begin{table*}
\begin{center}
  \caption{Priors and posterior median and confidence intervals for the different models \label{tab:parsmodels}}  
  \begin{tabular}{lcccc}
  \hline
  \hline
  \noalign{\smallskip}
   &  & Model 1 & Model 2 & Model 3 \\
  Parameter & Prior$^{(\mathrm{a})}$ & Posterior value$^{(\mathrm{b})}$
  & Posterior value$^{(\mathrm{b})}$ & Posterior value$^{(\mathrm{b})}$\\
  \noalign{\smallskip}
  \hline
  \noalign{\smallskip}
  \multicolumn{3}{l}{\emph{\bf \targetb's parameters }} \\
  \noalign{\smallskip}
    Orbital period $P_{\mathrm{orb}}$ (days)  & $\mathcal{N}[8.463000,0.000002]$ & \Pbdrs[] & \Pbbis[]  &\Pb[] \\
    Transit epoch $T_0$ (BJD$_\mathrm{TDB}-$2\,450\,000)  & $\mathcal{N}[8330.39051 , 0.00015]$ & \Tzerobdrs[] & \Tzerobbis[] & \Tzerob[]  \\  
    Eccentricity $e$  & $\mathcal{B}[1.52,29]^{(\mathrm{c})}$ & \ebdrs[]  & \ebbis[] & \eb[]  \\
    Angle of periastron $\omega$ (deg) &  $\mathcal{U}[0,360]$ & \wcdrs[]  & \wbbis[] & \wb[]  \\
    Doppler semi-amplitude variation $K$ (m s$^{-1}$) & $\mathcal{U}[0,50]$ & \kbdrs[]  & \kbbis[] & \kb[] \\
    \multicolumn{3}{l}{\emph{ \bf \targetc's parameters}} \\
    Orbital period $P_{\mathrm{orb}}$ (days)  & $\mathcal{N}[18.859019,0.000016]$ & \Pcdrs[]  & \Pcbis[] & \Pc[] \\
    Transit epoch $T_0$ (BJD$_\mathrm{TDB}-$2\,450\,000)  & $\mathcal{N}[8342.2223,0.0005]$ & \Tzerocdrs[]  & \Tzerocbis[] & \Tzeroc[]  \\ 
    Eccentricity $e$  & $\mathcal{B}[1.52,29]^{(\mathrm{c})}$ & \ecdrs[] & \ecbis[] & \ec[]  \\
    Angle of periastron $\omega$ (deg) &  $\mathcal{U}[0,360]$ & \wcdrs[]  & \wcbis[] & \wc[]  \\
    Doppler semi-amplitude variation $K$ (m s$^{-1}$) & $\mathcal{U}[0,50]$ & \kcdrs[]  & \kcbis[] & \kc[] \\
    \multicolumn{3}{l}{\emph{ \bf Multidimensional GP parameters}} \\
    GP Period $P_{\rm GP}$ (days) &  $\mathcal{U}[4.8,4.9]$ &  \jPGPdrs[]  & \jPGPbis[] & \jPGP[] \\
    $\lambda_{\rm p}$ &  $\mathcal{U}[0.1,2]$ & \jlambdapdrs[]  & \jlambdapbis[] &  \jlambdap[] \\
    $\lambda_{\rm e}$ (days) &  $\mathcal{U}[10,500]$ & \jlambdaedrs[]  & \jlambdaebis[] & \jlambdae[] \\
    $A_{\rm V}$ (\ms)  &  $\mathcal{U}[0,100]$ & \jAzerodrs[]  & \jAzerobis[] & \jAzero \\
    $B_{\rm V}$ (\ms\,d) &  $\mathcal{U}[-1000,1000]$ & \jAonedrs[]   & \jAonebis[] & \jAone \\
    $A_{\rm DLW}$  (1000 \mmss)  & $\mathcal{U}[-1000,1000]$ & \ldots   & \jAtwobis[] & \jAtwo \\
    $A_{\rm FWHM}$ (\kms)  & $\mathcal{U}[-1,1]$ & \jAtwodrs   & \ldots & \ldots \\
    $A_{\rm BIS}$ (\kms) &  $\mathcal{U}[-1,1]$ & \jAfourdrs   & \jAfourbis & \ldots \\
    $B_{\rm BIS}$ (\kms\,d) &  $\mathcal{U}[-1,1]$ & \jAfivedrs   & \jAfivebis & \ldots \\
    \multicolumn{3}{l}{\emph{ \bf Other parameters}} \\
    Offset  RV (SERVAL, \kms) & $\mathcal{U}[-1,1]$ & \ldots  & \RVsbis[] & \RVs[] \\
    Offset  RV (DRS, \kms) & $\mathcal{U}[-5,-3]$ & \RVsdrs[]  & \ldots & \ldots \\
    Offset  FWHM (\kms)  & $\mathcal{U}[10,12]$ & \FWHMdrs[]  & \ldots &  \ldots  \\
    Offset  DLW  (\mmss) & $\mathcal{U}[-1,1]$ & \ldots  & \DLWbis[] & \DLW[]  \\
    Offset  BIS (\kms) & $\mathcal{U}[-1,1]$ & \BISdrs[]  & \BISbis[] & \ldots  \\
    Jitter term $\sigma_{j,{\rm RV}}$ (\ms) & $\mathcal{J}[0,1000]$ & \jRVsdrs[]  & \jRVsbis[] & \jRVs[] \\
    Jitter term $\sigma_{j,{\rm FWHM}}$ (\ms)  & $\mathcal{J}[0,1000]$ &  \jFWHMdrs[] & \ldots & \ldots \\
    Jitter term $\sigma_{j,{\rm DLW}}$  (1000 \mmss) & $\mathcal{J}[0,1000]$ &  \ldots & \jDLWbis[] & \jDLW[] \\
    Jitter term $\sigma_{j,{\rm BIS}}$ (\ms)  & $\mathcal{J}[0,1000]$ &  \jBISdsr[]  & \jBISbis[] & \ldots \\
    \noalign{\smallskip}
    \hline
  \end{tabular}
\end{center}
  \begin{tablenotes}
  \item \emph{Note} -- $^{(\mathrm{a})}$, $^{(\mathrm{b})}$  and 
   $^{(\mathrm{c})}$ as defined in Table~\ref{tab:pars}.
\end{tablenotes}
\end{table*}

\section{Model comparison for different planet configurations}
\label{sec:planetarysignals}

Table~\ref{tab:comparison} shows a model comparison for different runs including different combinations of the \target\ b and c planetary signals (we note that the activity model used corresponds to Model 3 in Appendix~\ref{sec:threeruns}).

\begin{table*}
\begin{center}
  \caption{Model comparison \label{tab:comparison}}  
  \begin{tabular}{lcccccc}
  \hline
  \hline
Model & $N_{\rm pars}$ & $K_b$ [\ms] & $K_c$ [\ms] & $\ln \mathcal{L}$ & AIC$^a$ & BIC$^b$ \\
    \hline
Planet b signal only & 15 & $6.1_{-2.8}^{+2.9}$ & \ldots & 336 & -641 & -595  \\
Planet c signal only & 15 & \ldots & $8.4 \pm 2.6 $ & 339 & -648 & -601  \\
Both planetary signals & 20 & \kb[] & \kc[] & 342 & -643 & -581  \\
\hline
  \end{tabular}
\end{center}
  \begin{tablenotes}
  \item \emph{Note} -- $^{(a)}$ Akaike Information Criterion (AIC $= 2 N_{\rm pars} - 2 \ln \mathcal{L}$). $^{(b)}$ Bayesian Information Criterion (BIC $= N_{\rm pars} \ln N_{\rm data } - 2 \ln \mathcal{L}$).
\end{tablenotes}
\end{table*}

\section{TTV predictions}
\label{sec:TTV_pred}

\begin{figure*}
    \centering
    \includegraphics[width=\textwidth]{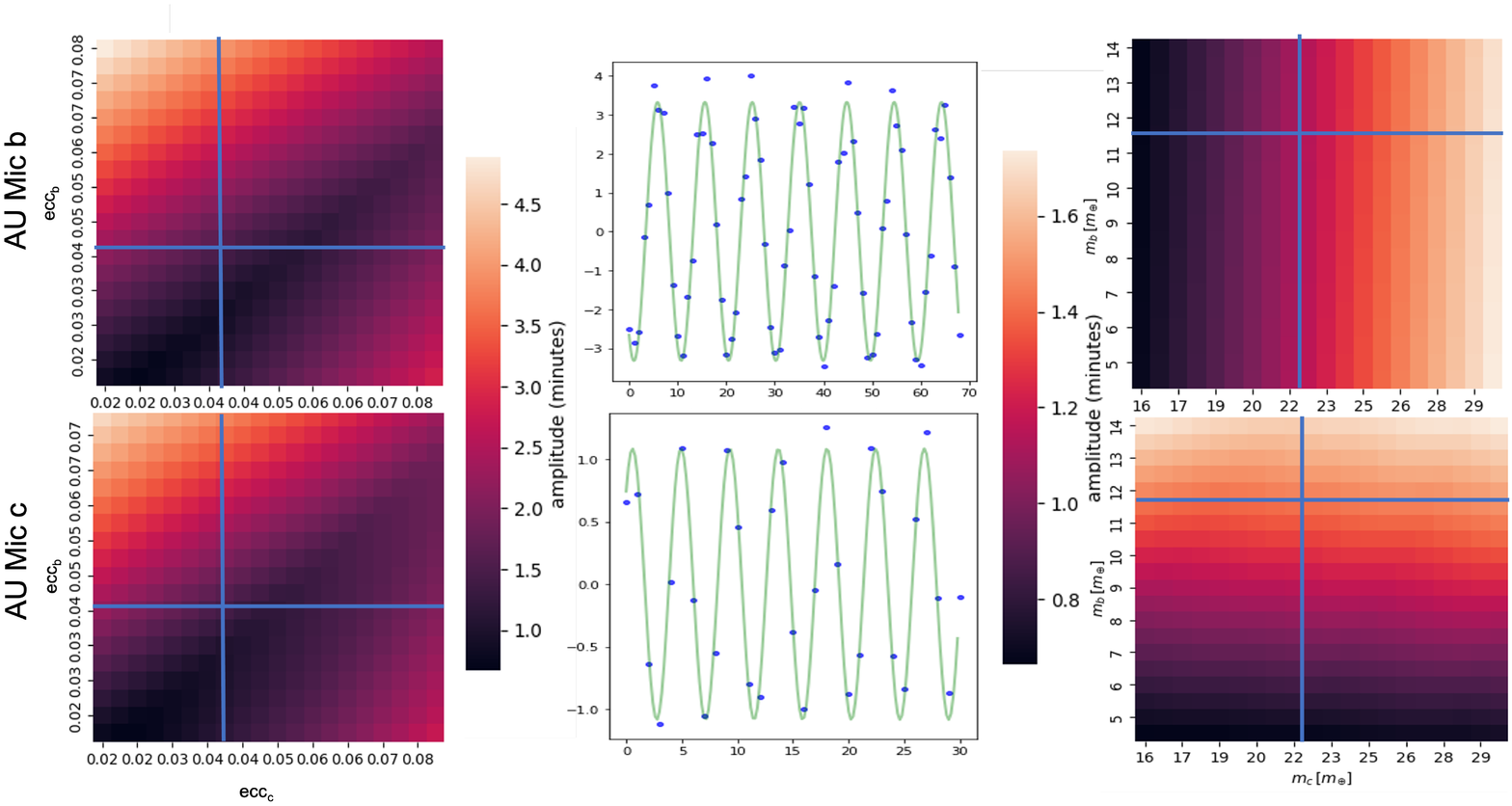}
    \caption{Expected TTV amplitudes for the \target\ system. In the left panel, we vary the eccentricities, while keeping the masses fixed. In the right panel, we keep the eccentricities fixed but vary the masses. In both panels, the vertical and horizontal solid lines indicate the maximum a posteriori values from our modeling of the RVs. Note the difference in colour scale between the left and right panels. The explored range of eccentricities and masses  extend slightly beyond the $1\sigma$ confidence limit in each direction. In the middle panel, the blue dots show the expected TTVs (time of transit centre relative to a linear ephemeris) when the planet parameters are set to their maximum a posteriori values. The x-axis corresponds to transit number. The green lines show a sinusoidal fit to those TTVs.}
    \label{fig:ttv_plot}
\end{figure*}

Figure~\ref{fig:ttv_plot} shows how the amplitude of the expected TTVs changes when we vary the planet masses or eccentricities, keeping the other parameters at the values reported in this work. Also shown are the TTVs expected for the nominal system parameters.

\bsp	
\label{lastpage}
\end{document}

%% file: aumic_params.tex
\newcommand{\sdistance}[1][pc] {$9.7248 \pm 0.0046 $ #1}
\newcommand{\smass}[1][\msun]{$0.50 _{- 0.03}^{ + 0.03} $ #1} 
\newcommand{\sradius}[1][\rsun]{$0.75 \pm 0.03 $ #1}
\newcommand{\stemp}[1][$\mathrm{K}$]{$3700 \pm 100 $ #1}
\newcommand{\slogg}{$4.39 \pm 0.03 $}
\newcommand{\slum}[1][\lsun]{$0.09 \pm 0.02 $ #1}
\newcommand{\sage}[1][Myr]{$22 \pm 3$ #1}
\newcommand{\srot}[1][days]{$4.86 \pm 0.01$ #1}
\newcommand{\siorb}[1][\degree]{$89.5 \pm 0.4$#1}
\newcommand{\svsini}[1][\kms]{$7.8 \pm 0.3$ #1}
\newcommand{\sllimb}{$0.2348$}
\newcommand{\sqlimb}{$0.3750$}

\newcommand{\Tzeroblit}[1][days]   {$8330.39051 \pm 0.00015$ #1}
\newcommand{\Tzeroclit}[1][days]   {$8342.2223 \pm 0.00015$ #1}
\newcommand{\impactb}[1][\rstar]{$0.18 \pm 0.11$ #1}
\newcommand{\impactc}[1][\rstar]{$0.51 \pm 0.21$ #1}
\newcommand{\pblit}[1][days]{$8.463000 \pm 0.000002$ #1}
\newcommand{\pclit}[1][days]{$18.859019 \pm 0.000016$ #1}
\newcommand{\semib}[1][au]{$0.0645 \pm 0.0013$ #1}
\newcommand{\semic}[1][au]{$0.1101 \pm 0.0022$ #1}
\newcommand{\radiusblit}[1][\rearth]{$4.07 \pm 0.17$ #1}
\newcommand{\radiusclit}[1][\rearth]{$3.24 \pm 0.16$ #1}
\newcommand{\massblit}[1][\mearth]{$17.1_{-4.5}{+4.7}$ #1}
\newcommand{\massclit}[1][\mearth]{$2.2 < M_{c} <25$ #1}
\newcommand{\kblit}[1][\ms]{$ 8.5 _{- 2.2.}^{ + 2.3}$ #1}
\newcommand{\kclit}[1][\ms]{$ 0.6 < K_{c} < 9.5$ #1}
\newcommand{\eqtempb}[1][K]{$ 593 \pm 21$ #1}
\newcommand{\eqtempc}[1][K]{$ 454 \pm 16$ #1}

%% file: model1_drs.tex
\newcommand{\Tzerobdrs}[1][days]   {$8330.39052 _{ - 0.00016 } ^ { + 0.00015 }$~#1} 
\newcommand{\Pbdrs}[1][days]   {$8.4629998 _{ - 2e-06 } ^ { + 2.1e-06 }$~#1} 
\newcommand{\ebdrs}[1][ ]   {$0.042 _{ - 0.027 } ^ { + 0.05 }$~#1} 
 
\newcommand{\kbdrs}[1][${\rm m\,s^{-1}}$]   {$5.89 _{ - 2.66 } ^ { + 2.98 }$~#1} 

\newcommand{\Tzerocdrs}[1][days]   {$8342.22227 _{ - 0.00046 } ^ { + 0.00048 }$~#1} 
\newcommand{\Pcdrs}[1][days]   {$18.859019 _{ - 1.6e-05 } ^ { + 1.5e-05 }$~#1} 
\newcommand{\ecdrs}[1][ ]   {$0.039 _{ - 0.025 } ^ { + 0.045 }$~#1} 
\newcommand{\wcdrs}[1][deg]   {$182.0 _{ - 126.0 } ^ { + 126.0 }$~#1} 
\newcommand{\kcdrs}[1][${\rm m\,s^{-1}}$]   {$6.7 _{ - 3.06 } ^ { + 3.09 }$~#1} 

\newcommand{\RVsdrs}[1][${\rm km\,s^{-1}}$]   {$-4.775 _{ - 0.013 } ^ { + 0.01 }$~#1} 
\newcommand{\FWHMdrs}[1][${\rm km\,s^{-1}}$]   {$10.7 _{ - 0.085 } ^ { + 0.074 }$~#1} 
\newcommand{\BISdrs}[1][${\rm km\,s^{-1}}$]   {$0.0068 _{ - 0.0073 } ^ { + 0.0076 }$~#1} 
\newcommand{\jRVsdrs}[1][${\rm m\,s^{-1}}$]   {$13.14 _{ - 1.82 } ^ { + 2.26 }$~#1} 
\newcommand{\jFWHMdrs}[1][${\rm m\,s^{-1}}$]   {$89.91 _{ - 7.0 } ^ { + 8.05 }$~#1} 
\newcommand{\jBISdsr}[1][${\rm m\,s^{-1}}$]   {$42.89 _{ - 3.52 } ^ { + 3.95 }$~#1} 
\newcommand{\jAzerodrs}[1][]   {$21 _{ - 14 } ^ { + 24 }$~#1} 
\newcommand{\jAonedrs}[1][]   {$102 _{ - 26 } ^ { + 53 }$~#1} 
\newcommand{\jAtwodrs}[1][]   {$0.149 _{ - 0.039 } ^ { + 0.076 }$~#1} 
\newcommand{\jAfourdrs}[1][]   {$-0.0048 _{ - 0.0119 } ^ { + 0.0097 }$~#1} 
\newcommand{\jAfivedrs}[1][]   {$-0.06 _{ - 0.032 } ^ { + 0.016 }$~#1} 
\newcommand{\jlambdaedrs}[1][]   {$142.7 _{ - 24.7 } ^ { + 35.1 }$~#1} 
\newcommand{\jlambdapdrs}[1][]   {$0.455 _{ - 0.048 } ^ { + 0.06 }$~#1} 
\newcommand{\jPGPdrs}[1][]   {$4.8604 _{ - 0.0031 } ^ { + 0.0032 }$~#1} 

%% file: model2_bis.tex
\newcommand{\Tzerobbis}[1][days]   {$8330.39051 _{ - 0.00015 } ^ { + 0.00016 }$~#1} 
\newcommand{\Pbbis}[1][days]   {$8.463 _{ - 2.1e-06 } ^ { + 2e-06 }$~#1} 
\newcommand{\ebbis}[1][ ]   {$0.042 _{ - 0.026 } ^ { + 0.044 }$~#1} 
\newcommand{\wbbis}[1][deg]   {$182.0 _{ - 129.0 } ^ { + 127.0 }$~#1} 
\newcommand{\kbbis}[1][${\rm m\,s^{-1}}$]   {$5.1 _{ - 2.15 } ^ { + 2.32 }$~#1} 
\newcommand{\Tzerocbis}[1][days]   {$8342.2223 _{ - 0.00051 } ^ { + 0.00051 }$~#1} 
\newcommand{\Pcbis}[1][days]   {$18.85902 _{ - 1.5e-05 } ^ { + 1.6e-05 }$~#1} 
\newcommand{\ecbis}[1][ ]   {$0.04 _{ - 0.026 } ^ { + 0.047 }$~#1} 
\newcommand{\wcbis}[1][deg]   {$143.3 _{ - 95.8 } ^ { + 141.4 }$~#1} 
\newcommand{\kcbis}[1][${\rm m\,s^{-1}}$]   {$8.86 _{ - 2.51 } ^ { + 2.42 }$~#1} 
\newcommand{\RVsbis}[1][${\rm km\,s^{-1}}$]   {$-0.0972 _{ - 0.005 } ^ { + 0.0077 }$~#1} 
\newcommand{\DLWbis}[1][${\rm km\,s^{-1}}$]   {$-0.001 _{ - 0.021 } ^ { + 0.024 }$~#1} 
\newcommand{\BISbis}[1][${\rm km\,s^{-1}}$]   {$0.0072 _{ - 0.007 } ^ { + 0.0063 }$~#1} 
\newcommand{\jRVsbis}[1][${\rm m\,s^{-1}}$]   {$10.32 _{ - 1.42 } ^ { + 1.68 }$~#1} 
\newcommand{\jDLWbis}[1][${\rm m\,s^{-1}}$]   {$30.29 _{ - 2.39 } ^ { + 2.78 }$~#1} 
\newcommand{\jBISbis}[1][${\rm m\,s^{-1}}$]   {$47.91 _{ - 3.73 } ^ { + 4.18 }$~#1} 
\newcommand{\jAzerobis}[1][]   {$10 _{ - 07 } ^ { + 15 }$~#1} 
\newcommand{\jAonebis}[1][]   {$102 _{ - 23 } ^ { + 39 }$~#1} 
\newcommand{\jAtwobis}[1][]   {$45 _{ - 10 } ^ { + 19 }$~#1} 
\newcommand{\jAfourbis}[1][]   {$-0.003 _{ - 0.0093 } ^ { + 0.008 }$~#1} 
\newcommand{\jAfivebis}[1][]   {$-0.05 _{ - 0.02 } ^ { + 0.012 }$~#1} 
\newcommand{\jlambdaebis}[1][]   {$111.7 _{ - 16.6 } ^ { + 17.3 }$~#1} 
\newcommand{\jlambdapbis}[1][]   {$0.46 _{ - 0.045 } ^ { + 0.054 }$~#1} 
\newcommand{\jPGPbis}[1][]   {$4.8572 _{ - 0.0025 } ^ { + 0.0037 }$~#1} 

%% file: aumic_pyaneti_params.tex
\newcommand{\Tzerob}[1][days]   {$8330.39051 \pm 0.00015 $~#1} 
\newcommand{\Pb}[1][days]   {$8.463000 \pm 0.000002 $~#1} 
\newcommand{\eb}[1][ ]   {$0.04 _{ - 0.025 } ^ { + 0.045 }$~#1} 
\newcommand{\wb}[1][deg]   {$179 _{ - 125 } ^ { + 128 }$~#1} 
\newcommand{\kb}[1][${\rm m\,s^{-1}}$]   {$5.8 \pm 2.5 $~#1} 
\newcommand{\mpb}[1][$M_{\oplus}$]   {$11.7 \pm  5.0 $~#1}

\newcommand{\Tzeroc}[1][days]   {$8342.22231 \pm 0.00050 $~#1} 
\newcommand{\Pc}[1][days]   {$18.859019 \pm 0.0000016 $~#1} 
\newcommand{\ec}[1][ ]   {$0.041 _{ - 0.026 } ^ { + 0.047 }$~#1} 
\newcommand{\wc}[1][deg]   {$153 _{ - 94 } ^ { + 124 }$~#1} 
\newcommand{\kc}[1][${\rm m\,s^{-1}}$]   {$8.5 \pm 2.5 $~#1} 
\newcommand{\mpc}[1][$M_{\oplus}$]   {$22.2 \pm 6.7 $~#1}

\newcommand{\RVs}[1][${\rm km\,s^{-1}}$]   {$-0.0982 _{ - 0.0043 } ^ { + 0.0047 }$~#1} 
\newcommand{\DLW}[1][${\rm km\,s^{-1}}$]   {$-0.003 _{ - 0.019 } ^ { + 0.018 }$~#1} 
\newcommand{\jRVs}[1][${\rm m\,s^{-1}}$]   {$10.3 _{ - 1.4 } ^ { + 1.6 }$~#1} 
\newcommand{\jDLW}[1][${\rm m\,s^{-1}}$]   {$30.7 _{ - 2.4 } ^ { + 2.7 }$~#1} 
\newcommand{\jAzero}[1][]   {$7.5 _{ - 05.5 } ^ { + 10.9 }$~#1} 
\newcommand{\jAone}[1][]   {$91 _{ - 18 } ^ { + 27 }$~#1} 
\newcommand{\jAtwo}[1][]   {$38.7 _{ - 08.5 } ^ { + 12.3 }$~#1} \newcommand{\jlambdae}[1][]   {$108 \pm 15 $~#1} 
\newcommand{\jlambdap}[1][]   {$0.449 _{ - 0.043 } ^ { + 0.049 }$~#1} 
\newcommand{\jPGP}[1][]   {$4.8571 _{ - 0.0027 } ^ { + 0.0037 }$~#1}
\newcommand{\jPGPtext}[1][]   {$4.857 _{ - 0.003 } ^ { + 0.004 }$~#1} 

\newcommand{\pdenb}[1][${\rm g\,cm^{-3}}$]{$0.97 \pm 0.43$}
\newcommand{\pdenc}[1][${\rm g\,cm^{-3}}$]{$3.66 \pm 1.28$}

%% file: RV_table.tex
\begin{table*}
\begin{center}
  \caption{Radial velocity observations and activity indicator  time-series. Observations which were affected by flares or clouds are flagged in the `Remarks' column. The full version of this table is available in machine-readable format as part of the supplementary material. \label{tab:rvs}}  
  \begin{tabular}{ccccccccccl}
  \hline
  \hline
  \noalign{\smallskip}
Time & RV$_{\rm SERVAL}$ & $\sigma_{\rm RV, SERVAL}$ & RV$_{\rm TERRA}$ &
$\sigma_{\rm RV, TERRA}$ & DLW & $\sigma_{\rm DLW}$ & \sshk\ & $\sigma_{\rm S_{HK}}$ & SNR  & Remarks \\
${\rm BJD_{TDB}} - 2\,450\,000$ & \kms & \kms & \kms & \kms & 1000 \mmss\ & & & & $\rm @550\,nm$  \\  \noalign{\smallskip}
  \hline
  \noalign{\smallskip}
7223.643264 & 0.1360 & 0.0060 & 0.2258 & 0.0059 & -12.2629 & 5.5325 & 7.8457 & 0.0237 & 71.2  \\ 
7223.652882 & 0.1333 & 0.0058 & 0.2252 & 0.0058 & -3.2790 & 4.8515 & 7.7974 & 0.0238 & 70.6 \\ 
7333.530396 & 0.0065 & 0.0024 & 0.1057 & 0.0023 & -40.2712 & 4.7079 & 7.8161 & 0.0160 & 129.1 \\ 
7333.541066 & 0.0110 & 0.0024 & 0.1066 & 0.0025 & -41.6257 & 4.8444 & 7.7178 & 0.0168 & 122.3 \\ 
7493.886516 & -0.0004 & 0.0040 & 0.1059 & 0.0048 & 43.9644 & 3.3771 & 8.8909 & 0.0134 & 146.9 \\ 
\ldots & \ldots & \ldots & \ldots & \ldots & \ldots & \ldots & \ldots & \ldots & \ldots \\ 
  \hline
  \end{tabular}
\end{center}
\end{table*}